\documentclass[a4paper,12pt]{article}

\usepackage{amsmath}
\usepackage{amsfonts}
\usepackage{amssymb}
\usepackage{simplewick}
\usepackage{a4wide}
\usepackage{booktabs}
\usepackage{xcolor}
\usepackage{feynmp-auto}
\def\Dirac{{\raise0.09em\hbox{/}}\kern-0.69em D}
\def\ep{i\epsilon} 

\def\kbar{{\mathchar'26\mkern-9muk}} 		
\def\lesssim{\mathrel{\hbox{\rlap{\hbox{\lower4pt\hbox{$\sim$}}}\hbox{$<$}}}}
\def\sq{\hbox{\rlap{$\sqcap$}$\sqcup$}}         
\def\p{\partial}                                
\def\tfrac #1#2{\textstyle{\frac{#1}{#2}}} 	
\def\dfrac #1#2{\displaystyle{\frac{#1}{#2}}} 	
\def\tr{\mbox{Tr}\,}                            

\def\beg{\begin{eg}\rm}                         
\def\eeg{\hfill\sq\end{eg}}                     

\def\Tr{{\rm Tr\,}} 

 \def\cS{{\cal S}}

\def\k {\kern-.1em\mathbin{,}\kern-.1em}
\def\hk{\kern.12em\raise-1em\hbox{$\hat{\raise1em\hbox{,}}$}\kern.12em}

\newcounter{eg}                                 
\newtheorem{eg}{Example}[section]
\def\beg{\begin{eg}\rm}                         
\def\eeg{\hfill\sq\end{eg}}                     
        
\newcommand{\initiate}{\setcounter{equation}{0}}        

\allowdisplaybreaks

\def\tA{{\sf A}}

\def\tF{{\sf F}}

\def\tX{{\sf X}}

\def\kbar{{\mathchar'26\mkern-9muk}}
\def\ep{i\epsilon}

\def\dif{d}
\def\tg{{\sf g}}

\begin{document} 
\unitlength = 1mm

\title{One-loop structure of the $U(1)$
 gauge model \\
on the truncated Heisenberg space} 
                        \vskip25pt
\author{Maja Buri\'c, 
Luka Nenadovi\' c
and         Dragan Prekrat\footnote{majab@ipb.ac.rs, lnenadovic@ipb.ac.rs, dprekrat@ipb.ac.rs}
                   \\
        University of Belgrade,  Faculty of Physics, P.O. Box 44
                   \\
        SR-11001 Belgrade 
       }
\maketitle
\parskip 10pt plus2pt minus2pt
%

\abstract{We calculate divergent one-loop corrections
to the propagators of the $U(1)$ gauge theory
on the truncated Heisenberg space,
which is one of the extensions of the Grosse-Wulkenhaar model.
The model is purely geometric, based on the Yang-Mills action; 
the corresponding gauge-fixed theory is BRST invariant. 
We  quantize  perturbatively and,
along with the usual wave-function and mass
 renormalizations, we find divergent nonlocal  terms
of the $\, \Box^{-1} \, $ and $\, \Box^{-2} \, $ type. 
We discuss the meaning of these terms and possible 
improvements of the model. }

\setlength{\parskip}{5pt plus2pt minus2pt}

\thispagestyle{empty} 

\initiate

\section{Introduction}

Unsuccessful attempts to quantize  gravitational field
by  usual methods indicate that the 
structure of spacetime at the Planck scale
 is very different from the classical one.
Such conclusion is also indicated by the existence 
of singularities in general relativity and divergences in 
perturbative quantum field theory.
There are many reasons to believe that the theory of 
quantum gravity will not be local in the conventional sense,
 \cite{Douglas:2001ba}; however, 
 locality is in the core of standard field theories
and one has to solve many problems
in order  to formulate a consistent framework to describe 
nonlocal classical and quantum fields. 

Perhaps the first idea of how to `delocalize'
 points was the Kaluza-Klein extension of
spacetime  by additional compact  dimensions. 
Alternatively, nonlocality can be introduced
through the structure of elementary constituents as in string theory;
in both cases the underlying spacetime  
is a Riemannian manifold. 
Another way to introduce nonlocality is to assume that 
spacetime is described by an algebra of noncommuting operators. 
We analyze this approach, presuming
that noncommutativity of coordinates 
of `quantum spacetime' is physical;
we further investigate properties of quantum fields
defined on it. Our motivation for this study is twofold. 
The first  is the expectation that the existing knowledge about
operator algebras and their representations gives 
 enough tools to build a compact mathematical framework. The  second
is the hope that the `amount of nonlocality' introduced
algebraically is  restricted and that it can
provide with reasonable and interesting physics.

Mathematical and physical
results which concern noncommutative geometry and
noncommutative field theories are numerous. In a brief summary
 one can say that geometry and classical field theories are 
understood fairly well, whereas quantization is still an open 
problem. A technical explanation is that, though 
in quantization the effects of nonlocality improve
the ultraviolet behavior of a theory, 
the mixing of large and small length scales induced by
noncommutativity transfers divergences to the infrared sector
(UV/IR mixing). This is clearly established for theories defined 
on the Moyal space. On the other hand, quantum field theories defined 
on spaces with finite matrix representations  are finite, and thus 
they can be viewed as matrix regularizations of (the corresponding)
commutative  theories. Moreover there are models in which 
the continuous limit of a matrix theory has better quantization properties 
than its commutative  predecessor, \cite{Chu:2001xi}. 

First fully  renormalizable 
theory formulated on a noncommutative space is
the Grosse-Wulkenhaar (GW) model \cite{Grosse:2003nw,Grosse:2004yu}.
It describes  real scalar field on the Moyal space
of Euclidean signature confined in the oscillator potential. 
The potential term induces symmetry  between  long and short distances,
the so-called Langman-Szabo (LS) duality \cite{Langmann:2002cc}, 
which is roughly of the form $\, x\leftrightarrow p$. 
The  field propagator is given by the Mehler kernel:  it
regularizes the UV behavior of the theory keeping 
the IR sector finite. Renormalizability of the GW model is
 established and thoroughly examined  by various methods
\cite{Gurau:2005gd,Rivasseau:2006,Grosse:2004by,Disertori:2006nq};
its exceptional mathematical properties  induced a lot of subsequent work.
Two renormalizable spinor field theories 
analogous to the GW model have been found: in the first one \cite{Grosse:2007jy}, 
the representation of spinors is chosen in
such  way that the square of the propagator is the Mehler kernel.
The other model, of Vignes-Tourneret (VT) \cite{VignesTourneret:2006nb},
 is a noncommutative generalization of the Gross-Neveu model 
\cite{Gross:1974jv}. In both theories the Lagrangian has an explicit dependence on 
coordinates and breaks the translation invariance. To improve the last
property  an interesting translationally invariant model, which
is also renormalizable,  was proposed in \cite{Gurau:2009}. 
It contains, instead of the oscillator potential, the $\Box^{-1}$ term  
in the kinetic part of the action, thus introducing another 
version  of the LS duality:   $\,\Box \leftrightarrow \Box^{-1}\,$. 

All attempts to find renormalizable gauge model { \`a la} Grosse-Wulkenhaar 
have been by now unsuccessful. Several strategies have been used, 
all mainly based on parallels  with
standard gauge theories. As most of the results are thoroughly 
reviewed in \cite{Blaschke:2010kw} we will here   
recall only some guiding ideas. The first proposal
was to transform the gauge propagator into the  Mehler kernel
by nonlinear gauge fixing \cite{Blaschke:2007vc}; 
however, the appearance of the tadpole divergence
made the theory nonrenormalizable. 
Another strategy was to  impose  the LS duality
\cite{de Goursac:2007gq,Grosse:2007dm}: the action for the gauge 
field was defined by minimal coupling to the GW scalar 
and subsequent  integration of the scalar field. Although the
obtained, induced theory has good symmetry properties
(the LS duality becomes the invariance under exchange
$[x^\mu,\ ]\leftrightarrow \{x^\mu,\ \} $), it 
does not have the trivial vacuum solution and
perturbative quantization is not well defined. It is
important to notice  that the explicit coordinate 
dependence of the induced gauge theory can be  elegantly
rewritten using the covariant coordinates which were
introduced much earlier, \cite{Madore:2000en}. For recent  results  
on quantization of this theory in the matrix base
we refer to \cite{Martinetti:2013uia}.

The idea which we have developed in the previous papers is that 
specific forms of the GW and  VT actions are due to the underlying 
(noncommutative) geometry. The idea is based on the result
that the two-dimensional GW action can be viewed as 
an action defined on  particular curved 
three-dimensional space after the Kaluza-Klein (KK) reduction, 
\cite{Buric:2009ss}. The VT action, similarly,  is the  spinor 
action on the same space, \cite{Buric:2015vja}.
In both cases matter is nonminimally   coupled
to the background curvature and torsion. The employed
approach gives also the
$U(1)$ Yang-Mills (YM) theory which consists,
after the KK reduction, of interacting gauge and scalar
fields. Classical properties of the model are
very good:  there are vacuum solutions 
which include the trivial vacuum, the BRST invariance 
is established, \cite{Buric:2010xs}. The perturbative quantization  
was started in \cite{Buric:2012bb}  with the calculation of 
 divergences of the first order in the gauge coupling:
the obtained divergences were IR logarithmic
and included the  tadpole. It was however hard to systematize the 
computation of the  prefactors. The 2-point divergences
which were found can be removed by the usual mass and wave 
function renormalizations, but the tadpole diagram remains, signalling
instability of the trivial vacuum under quantum fluctuations.

We continue here investigation of the quantization properties
of the proposed gauge model. We calculate  one-loop corrections 
to the propagators of  second order;  we find a systematic
way to compute divergent integrals with two or more parameter
integrations, which enables us to compare and 
add various contributions. However, 
in addition to the local terms, we find new `nonlocal' 
infrared divergences of the $\Box^{-1}\,$ and
the $\Box^{-2}\,$ type. Such terms do not exist in
the classical action, thus rendering the theory 
nonrenormalizable.

The paper is organized as follows. In Section~2 we define 
and briefly review properties of the truncated Heisenberg space
and the Yang-Mills theory on it, recollecting  results from 
\cite{Buric:2010xs}. In Section~3 we go through the main steps
and  some details of the calculation and list 
additional propagator corrections, completing the earlier result
 \cite{Buric:2012bb}. In Section~4 we discuss the meaning of
the obtained results and possibilities to improve the model. 
Important details of  calculation are given in the Appendices.

\initiate

\section{Fields on the truncated Heisenberg space}

Truncated Heisenberg space is a noncommutative space 
 $\cal{A}$  generated by three hermitian coordinates $\, x$, $y$, $z$
 which satisfy commutation relations:
\begin{eqnarray}
 [ x, y] = i \epsilon\mu^{-2} \,(1-\mu z) , 
 \quad [ x, z ]  = i\epsilon \,( y  z+ z  y)  , 
 \quad      \label{alg}
 [y, z ] = - i\epsilon \,( x  z+  z x) .     
\end{eqnarray} 
Constant $\mu\,$ has dimension of the inverse length and
  $\epsilon$ is a dimensionless noncommutativity parameter.
For $\epsilon=1$ algebra (\ref{alg}) has finite-dimensional 
matrix representations; 
$\,\epsilon=0\,$  defines the `commutative limit'. 
Double scaling limit $\mu \to 0$, $\epsilon\to 0$,
 $ \, \kbar = \epsilon \mu^{-2}=$ finite reduces (\ref{alg}) to 
the Heisenberg algebra
\begin{equation}
\qquad\ 
 [x,y] =i\kbar .                                     \label{algH}
\end{equation}
Irreducible representation of the Heisenberg algabra
 is infinite-dimensional;  in the
 geometric context it is called the Moyal plane.  Truncation of 
 infinite matrices $x$, $y$, $z$ given in the Fock representation to
finite $n\times n$ matrices gives algebra (\ref{alg}).
In this sense  Heisenberg algebra (\ref{algH}) is a contraction, 
or subspace $z=0$,  of the truncated Heisenberg space 
\cite{Buric:2011xb}. Limit $n\to\infty$ which transforms 
(\ref{alg}) to (\ref{algH}) is a weak operator limit.

The truncated Heisenberg algebra can be endowed with 
differential structure. The space 
of 1-forms is spanned by  frame  $\{ \theta^\alpha\} $,
$\alpha = 1,2,3$; derivations $e_\beta$ dual to 
 $\theta^\alpha$ are defined to satisfy
$\,\theta^\alpha(e_\beta) =\delta^\alpha_\beta$.
We  assume, \cite{book}
\begin{equation}
 [f,\theta^\alpha] =0 ,\qquad 
df =( e_\alpha f) \,\theta^\alpha =  [p_\alpha, f] \,\theta^\alpha .       
\label{df}
\end{equation}
The frame derivations $e_\alpha$ are inner and generated by 
momenta $p_\alpha\in\cal{A}$; $p_\alpha$ are, by convention, antihermitian. 
An important property of the inner-derivation calculus
 is the existence of a special connection 
$$  \theta = -p_\alpha \theta^\alpha  $$
which generates the differential, $ \, df = -[\theta,f]. $
We choose
\begin{equation}
 \epsilon p_1 =i \mu^2 y, \qquad \epsilon p_2 =-i\mu^2 x,\qquad 
\epsilon p_3 =i\mu \left(\mu z- \frac 12\right) .
\end{equation} 
It can be easily seen  that for
 $z=0\,$ this differential reduces to the
standard one on the Moyal plane.

The algebra of  momenta is in general quadratic, \cite{book}
\begin{equation}
 2P^{\gamma\delta}{}_{\alpha\beta}p_\gamma p_\delta 
- F^\gamma{}_{\alpha\beta} p_\gamma -\frac{1}{\ep} K_{\alpha\beta} =0, 
\label{2*}
\end{equation}
the $K_{\alpha\beta}  $, $F^\gamma{}_{\alpha\beta} $ and  
$ P^{\gamma\delta}{}_{\alpha\beta}$ are constants.
It defines a  noncommutative wedge product. 
The Hodge dual on the other hand cannot be defined in the general case as
 it depends on (the existence of) trace: in our case it is almost 
unique,  \cite{Buric:2015vja}. Finally, one  specifies the connection:
the metric-compatible connection used in \cite{Buric:2009ss}
defines a noncommutative space with curvature and torsion. 

The $U(1)$ gauge symmetry is introduced through the
gauge potential $\tA$  which is an antihermitian 1-form,  and the 
field strength $\tF$:
\begin{equation}
 \tA = i\tg\tA_\alpha \theta^\alpha ,\qquad
 \tF = d\tA +\tA^2 = \frac i2 \, \tF_{\alpha\beta} \theta^\alpha \theta^\beta .
\end{equation} 
The $\tg$ denotes the $U(1)$ coupling constant;
the $U(1)$  group consists of all unitary elements of $\cal{A}$. 
A remarkable property of noncommutative differential which
we use is a possibility to construct a  gauge-covariant 1-form: the difference 
\begin{equation}
 \tX =\tX_\alpha \theta^\alpha = \tA-\theta ,\qquad \tX_\alpha = p_\alpha + i\tg \tA_\alpha 
\end{equation}
transforms in the adjoint representation of the gauge group. 
Coefficients $ \tX_\alpha$ are called  covariant coordinates (a more  
appropriate name would   perhaps be covariant momenta).
Expressing the field strength in terms of $\, \tX$ and the 
structure constants we find
\begin{eqnarray}
 && \tF = \tX^2  - \frac 12\, F^\gamma{}_{\alpha\beta}{\tX}_\gamma 
 \,\theta^\alpha\theta^\beta -\frac{1}{2i\epsilon}\, K_{\alpha\beta}\, 
 \theta^\alpha\theta^\beta     . \label{FX} 
\label{Falbe} 
\end{eqnarray}
The existence of $\,\tX$ means that there are  covariant observables
which depend only on the potentials, and it
opens a possibility to define alternative actions
for gauge fields, with different properties from
the Yang-Mills or the Chern-Simons actions.
This is a new effect characteristic for noncommutative spaces.
Our model is, however, built as a  
noncommutative generalization of the Yang-Mills theory so we shall
keep only the original YM term in the action;
we  discuss possible new terms in the last section. 

The YM action  on the truncated Heisenberg space is given by
\begin{equation}
 \cS_{YM} = \frac{1}{16 \tg^2}\, \Tr\big(\tF(*\tF) +(*\tF)\tF\big). \label{210}
\end{equation} 
Dimensional reduction to  $z=0\,$ is done by considering only fields
 $\,\tA_\alpha(x,y,z=0)$, by (formally) integrating 
 over $z$ and by rescaling  gauge coupling constant ($\tg\to g$)
 and  gauge fields. 
 This gives the Kaluza-Klein reduced  action on the  Moyal plane. 
 In order to distinguish the values  of  gauge fields 
 $\,\tA_\alpha$, $\tF_{\alpha\beta}\, $, $\alpha,\beta =1,2,3$\, 
defined in three dimensions from the gauge fields 
defined intrinsically on the  Moyal plane, we denote the latter by 
$A_\alpha, \ F_{\alpha\beta}$, $\alpha,\beta = 1,2$. Fields and
coupling constants have different mass dimension in two and three 
dimensions: dimensional reduction procedure takes care of this 
automatically. 
For $z=0\,$ the third component of the momentum is constant,
$\,p_3 =-{i\mu}/{2\epsilon}\, $, $\,e_3=0$, and
 $\tA_3$ transforms as a
scalar field in the adjoint representation. We  denote 
\begin{eqnarray}
 && \tg \tA_3 = g \phi,\ \qquad \tg \tA_1 =  g A_1,\qquad
 \tg \tA_2= g A_2 .
 \end{eqnarray}
The field strength and  covariant derivative 
 in two dimensions  are defined as
 \begin{eqnarray}
 && {D}_\alpha \phi = e_\alpha \phi+  
ig[A_\alpha, \phi] ,\quad
g^{-1} {F_{12}} = e_1A_2 -e_2 A_1 + ig[A_1, A_2].
\end{eqnarray}
After the KK-reduction, components of the three-dimensional  $\tF$ become, 
\cite{Buric:2010xs}
\begin{eqnarray}
&&\tg^{-1}\tF_{12} = g^{-1} F_{12}-\mu  \phi = g^{-1} \! \left(-i [\tX_1,\tX_2] + 
\frac{\mu^2}{\epsilon}\right)-\mu \phi ,           
\nonumber  \\[6pt]
&&\tg^{-1}\tF_{13} ={D}_1\phi - i\epsilon  \{ p_2 + igA_2, \phi\} = [\tX_1, \phi] 
-i\epsilon \{ \tX_2, \phi\} , \label{field} \\[8pt]
&&\tg^{-1}\tF_{23} = {D}_2\phi + i\epsilon  \{ p_1 +igA_1, \phi\} = [\tX_2,\phi] +i\epsilon 
\{ \tX_1, \phi \}\nonumber .
\end{eqnarray}
Introducing
\begin{equation}
 a=1-\epsilon^2
\end{equation}
 we obtain
\begin{equation}
 \cS_{YM} =  \frac{1}{2 \tg^2}\, \tr \, \big( a\, \tF_{12}\tF^{12} +
\tF_{13}\tF^{13} +\tF_{23}\tF^{23}\big)   ,
\end{equation} 
that is,
\begin{eqnarray}
 &&\cS_{YM}= \frac 12 \,\tr \Big( \frac{a}{g^2}\, 
 (F_{12})^2 -\frac{2 a\mu}{g}\, F_{12}\phi 
+ (4+a) \mu^2 \phi^2 - 4\epsilon F_{12}\phi^2    \label{L}
\\[0pt]
&&\phantom{S = \frac 12 \tr \quad\ } + (D_1 \phi)^2 +(D_2 \phi)^2
 -\epsilon^2\{ p_1 +igA_1, \phi\}^2 -\epsilon^2\{ p_2 +igA_2, \phi\}^2 \Big)
  \qquad  \nonumber  \\
 &&\phantom{\cS_{YM}}= \frac 12 \, \tr \Big(-\frac{a}{g^2} \, 
 [\tX_1,\tX_2]^2+ a\mu^2\phi^2-\frac{2a\mu^3}{g \epsilon}\,\phi
 +\frac{2ia\mu}{g}\,[\tX_1,\tX_2]\,\phi
\qquad \nonumber \\
&&\phantom{\cS_{YM}= \tfrac 12 \tr  \ }
+ 4\ep \,[\tX_1,\tX_2]\,\phi^2 +[\tX_1,\phi]^2 +[\tX_2,\phi]^2 
-\epsilon^2\{ \tX_1,\phi\}^2
-\epsilon^2\{ \tX_2,\phi\}^2  \Big)\nonumber .
\end{eqnarray} 
Two expressions are the same up to terms which are constant or
proportional to a commutator, that is, to  surface and cosmological
constant terms\footnote{The background noncommutative space is curved
but gravity is not dynamical.}.

Let us briefly analyze actions (\ref{210}) and (\ref{L}). 
Clearly, they are defined  only when the trace is defined, 
that is in a fixed representation of the algebra. One way to proceed
 is to consider finite matrix representations,
that is $\epsilon =1$, $a=0$: it gives a
 non-propagating gauge field which interacts
with the scalar. Another possibility, which we 
choose here, is to go to the continuous limit 
and represent fields on the Moyal space. There are
various advantages and drawbacks of this choice. On the
one hand, the resulting action is relatively complicated
as gauge and  scalar fields are mixed
in the kinetic  term. This fact
on the other hand  indicates that the harmonic
potential  confines both fields,  gauge and scalar. The action is
manifestly gauge invariant, but the status of the LS duality 
is not clear: (\ref{L}) is not invariant under
the exchange $\, [\tX_1,\tX_2]\leftrightarrow \{\tX_1,\tX_2\}\,$.
However, one hopes that the geometric origin of the action could
induce  cancellation of divergences as in supersymmetry.

Action (\ref{L}) has two classical vacua,
\begin{eqnarray}
 &&  A_1 =0,\quad A_2 =0,\quad \phi =0,
\\[4pt]
 &&  A_1 =- \frac{\mu^2 y}{g\epsilon} \, 
 ,\quad A_2 =\frac{\mu^2 x}{g\epsilon}\, ,
\quad \phi =\frac{\mu}{g\epsilon}\, .          \label{vac}
\end{eqnarray} 
The first is the usual trivial vacuum; the second describes 
a configuration with  constant value of the field strength
$F_{12} = {\mu^2}/{\epsilon}$. In quantization
we expand around the trivial vacuum. After the gauge fixing 
and inclusion of the ghost terms, we obtain,
\cite{Buric:2012bb}
\begin{equation}
  S=S_{YM}+S_{gf}+S_{gh}=S_{kin}+S_{int},
\end{equation}
with
 \begin{align}
  S_{kin}=&-\frac{1}{2}\int aA_\alpha\square A^\alpha
+2a\mu\epsilon^{\alpha\beta}(\partial_a A_\beta)
  \phi+\phi\square\phi  \label{Kin} \\
  &\qquad -(4+a)\, \mu^2\phi^2-4\mu^4 x^\alpha
 x_\alpha\phi^2+2\bar{c}\square c ,
 \nonumber\\[10pt]
  S_{int}=&-\frac{1}{2}\int 4\epsilon g\epsilon_{\alpha\beta}
(\partial^\alpha A^\beta+igA^\alpha
  \star A^\beta)\star\phi^2
  -2ig(\partial_\alpha\phi)[A^\alpha\stackrel{\star}{,}\phi]
 \label{Sint} \\[0pt]
  &\qquad +2ia\mu g\epsilon_{\alpha\beta}A^\alpha\star
  A^\beta\phi-2ia g\epsilon_{\alpha\beta}\partial^\alpha 
  A^\beta\epsilon_{\gamma\beta}A^\gamma\star A^\delta
  \nonumber
  \\[6pt]
  &\qquad +a g^2(\epsilon_{\alpha\beta}A^\alpha\star 
  A^\beta)^2+ g^2[A_\alpha\stackrel{\star}{,}\phi][A^\alpha
  \stackrel{\star}{,}\phi]
    -\epsilon^2 g^2\{A_\alpha\stackrel{\star}{,}\phi\}
\{A^\alpha\stackrel{\star}{,}\phi\}\nonumber\\[6pt]
  &\qquad +2\mu^2\epsilon g\epsilon_{\alpha\beta}
\{x^\alpha\stackrel{\star}{,}\phi\}
\{A^\beta\stackrel{\star}{,}\phi\}
  -i g\bar{c}\partial_\alpha[A^\alpha\stackrel{\star}{,}c] 
.\nonumber
  \end{align}  
This is the action which we will analyze. 

\initiate

\section{Propagators: the one-loop structure}

Let us recall some  results from \cite{Buric:2012bb} and
introduce new notation which enables us to perform
calculations more efficiently. We start with
the kinetic term. The scalar and  gauge fields 
in the kinetic term are mixed: because of noncommutativity  it is not possible 
to diagonalize it. We therefore consider fields as multiplet
 $\, \Phi^T=(A^\mu, \phi)$ (which they  were before the KK reduction), 
 and write the kinetic term as 
\begin{eqnarray}
 && S_{kin}=-\frac{1}{2}\int \left(\begin{array}{cc}
 A^\mu &\phi
\end{array} \right)
  \left(\begin{array}{cc}
         a\, \Box\delta_{\mu\nu}&-a\mu\epsilon_{\mu\xi}\partial^\xi \\[4pt]
         a\mu\epsilon_{\nu\eta}\partial^\eta& K^{-1}-a\mu^2
        \end{array}\right)
  \left(\begin{array}{c}
  A^\nu \\[2pt]
  \phi         \end{array}\right)
+2\bar{c}\,\Box c    \nonumber                \\[4pt]
&&\phantom{S_{kin}}
 = -\frac{1}{2}\int  \Phi^T G^{-1}\Phi +2\bar{c}\,\Box c
\label{Skin}
\end{eqnarray}
where we introduced
\begin{equation}
 K^{-1}=\Box-4\mu^4 x_\alpha x^\alpha-4\mu^2 .
\end{equation}
The  corresponding inverse operator, the momentum 
space Mehler kernel for the massive scalar field, has  in 
two dimensions  the followibg parametric form, \cite{kronb}
 \begin{equation}
{K}(r,s)=-\frac{\pi}{4\mu^4}
\int\limits^{\infty}_{1}
\frac{d\xi}{\xi}\frac{\xi-1}{\xi+1} \,
 e^{-\frac{1}{8\mu^2}\big((r+s)^2 \xi+(r-s)^2\frac{1}{\xi}\big)} .
 \end{equation}  
Mass of the scalar field is $2\mu$; for other values of mass,
factor $(\xi-1)/(\xi+1)$ has a different exponent.
We  denote
\begin{equation}
 \tilde{r}^\mu=\epsilon^{\mu\nu}r_\nu,
\qquad
r\wedge s=\frac{\epsilon}{\mu^2}\,\epsilon_{\mu\nu}r^\mu s^\nu
=\frac{\epsilon}{\mu^2}\, r\cdot \tilde s .
\end{equation}
The momentum-space kernel of the kinetic  operator   is
\begin{equation}
G^{-1}(r,s)=
\begin{pmatrix}
                     & -ar^2 \delta_{\mu\nu}  (2\pi)^2  \delta(r+s)
& -ia\mu\tilde{r}_\mu  (2\pi)^2 \delta(r+s)  &
                     \\ \\
                     &-ia \mu\tilde{s}_\nu (2\pi)^2 \delta(r+s)  
&-a\mu^2 (2\pi)^2 \delta(r+s)  +K^{-1}(r,s) 
                    \end{pmatrix}  . \label{G-1}
\end{equation}
Inverting it, for the matrix elements of the propagator $G(r,s)$ we obtain
\begin{align}
 & \bcontraction[1ex]{\,}{}{\phi(r)}{}\phi(r)\phi(s)=K(r,s),
 \nonumber\\[4pt] \label{contract}
& \bcontraction[1ex]{\,}{}{A^\alpha(r)}{}A^\alpha(r)\phi(s)=-i\mu\,
\frac{\tilde{r}^\alpha}{r^2}\, K(r,s), \\[4pt]
 &\bcontraction[1ex]{\,}{}{A^\alpha(r)}{}A^\alpha(r)A^\beta(s)=
(-i\mu)^2\, \frac{\tilde{r}^\alpha\tilde{s}^\beta}{r^2s^2}\, K(r,s)
			   -\frac{(2\pi)^2}{a}
\frac{\delta^{\alpha\beta}\delta(r+s)}{r^2} \, .\nonumber
\end{align}
These matrix elements obey the recurrence relations
\begin{align}
& \bcontraction[1ex]{\,}{}{A^\alpha(r)}{}A^\alpha(r)\phi(s)
=-i\mu\, \frac{\tilde{r}^\alpha}{r^2}\,
 \bcontraction[1ex]{\,}{}{\phi(r)}{}\phi(r)\phi(s), 
 \label{recur}\\[4pt]
& \bcontraction[1ex]{\,}{}{A^\alpha(r)}{}A^\alpha(r)A^\beta(s)
=(-i\mu)^2\, \frac{\tilde{r}^\alpha\tilde{s}^\beta}{r^2s^2}\,
 \bcontraction[1ex]{\,}{}{\phi(r)}{}\phi(r)\phi(s)		  
 -\frac{(2\pi)^2}{a}\frac{\delta^{\alpha\beta}\delta(r+s)}{r^2} 
\nonumber
\end{align}
which we will later use.

The interaction  contains three- and four-vertices.
In  momentum space they are
\begin{align}
S_{int,1}&=-\frac{2i\epsilon g}{(2\pi)^4}
\int dp\,dq\,dk\,
\delta(p+q+k)
\cos\frac{p\wedge q}{2} \,\tilde{p}^\mu 
A_\mu(p) \, \phi(q) \, \phi(k)	
\nonumber
\\ 
  S_{int,2}&=\frac{2ig}{(2\pi)^4}
\int dp\,dq\,dk\,
\delta(p+q+k)
\sin\frac{p\wedge q}{2}\, p^\mu 
\phi(p) \, \phi(q) \, A_\mu(k)	
\nonumber
\\
S_{int,3}&=-\frac{4i\epsilon\mu^2 g}{(2\pi)^4}
\int dp\,dq\,dk\,
\delta(p+q+k)
\cos\frac{p\wedge q}{2} \frac{\partial}{\partial\tilde{p}_\mu} 
\phi(p) \, \phi(q) \, A_\mu(k)	 
\nonumber
\\ 
S_{int,4}&=-\frac{a\mu g}{(2\pi)^4}
\int dp\,dq\,dk\,
\delta(p+q+k)
\sin\frac{p\wedge q}{2} \,\epsilon^{\mu\nu} 
A_\mu(p) \, A_\nu(q) \, \phi(k)  
\nonumber
\\ 
S_{int,5}&=\frac{ia g}{(2\pi)^4}
\int dp\,dq\,dk\,
\delta(p+q+k)
\sin\frac{p\wedge q}{2} \,\epsilon^{\mu\nu} \tilde{k}^\rho 
A_\mu(p) \, A_\nu(q) \, A_\rho(k) 
\nonumber
\\
S_{int,6}&=\frac{2ig}{(2\pi)^4}
\int dp\,dq\,dk\,
\delta(p+q+k)
\sin\frac{p\wedge q}{2}\, p^\mu \,
\bar{c}(p) \, c(q) \, A_\mu(k)
\nonumber\\
S_{int,7}&=\frac{2 g^2}{(2\pi)^6}
\int dp\,dq\,dp'\,dq'\,
\delta(p+q+p'+q')
\sin\frac{p\wedge q}{2} \sin\frac{p'\wedge q'}{2} \times 
\nonumber\\
& \hspace{.35\textwidth}\times\delta^{\mu\nu} 
A_\mu(p) \, \phi(q) \, A_\nu(p') \, \phi(q') 
\nonumber
\\ 
S_{int,8}&=\frac{2\epsilon^2 g^2}{(2\pi)^6} 
\int dp\,dq\,dp'\,dq'\,
\delta(p+q+p'+q')
\cos\frac{p\wedge q}{2} \cos\frac{p'\wedge q'}{2} \times
 \nonumber\\ & \hspace{.35\textwidth}\times \delta^{\mu\nu}
A_\mu(p) \, \phi(q) \, A_\nu(p') \, \phi(q') 
\nonumber
\\ 
S_{int,9}&=-\frac{2\epsilon g^2}{(2\pi)^6}
\int dp\,dq\,dp'\,dq'\,
\  d(p+q+p'+q')
\sin\frac{p\wedge q}{2} \cos\frac{p'\wedge q'}{2} \times
 \nonumber\\  &\hspace{.35\textwidth}\times\epsilon^{\mu\nu} 
A_\mu(p) \, A_\nu(q) \, \phi(p') \, \phi(q')	
\nonumber
\\ 
S_{int,10}&=\frac{a g^2}{2(2\pi)^6}
\int dp\,dq\,dp'\,dq'\,
\  d(p+q+p'+q')
\sin\frac{p\wedge q}{2} \sin\frac{p'\wedge q'}{2}\epsilon^{\mu\nu}\times
 \nonumber\\ &\hspace{.35\textwidth} \times \epsilon^{\rho\sigma}
A_\mu(p) \, A_\nu(q) \, A_\rho(p') \, A_\sigma(q').
\nonumber
\end{align}

We want to calculate the 
one-loop  corrections to the propagators, that is, sum 
of the expectation values 
\begin{align}
P_{FF',ij}(r,s)&=-\langle F(r)F'(s)\, S_{int,i} \,S_{int,j}\,\rangle  ,
\qquad
i,j=1,\ldots,6
\label{ij}
\\[4pt]
P_{FF',i}(r,s)&=-\langle F(r)F'(s)\, S_{int,i}\,\rangle , 
\qquad\  \qquad 
i=7,\ldots,10
\label{i}
\end{align}
where $F$ an $F'$ are fields  $\,\phi$ or $\,A^\mu$ and 
 $i,j$ label different interaction vertices, 1-10. Expressions 
of the form (\ref{ij}) correspond to the 2-point functions;
(\ref{i}) are the 1-point functions. We have previously calculated  
divergent 1-point functions $ P_{F,i}$,
 \cite{Buric:2012bb}. In fact, as we wish to obtain 
divergent terms as they  appear in the effective action, we can
go a step further and calculate the amputated graphs $\Pi(r,s)$. 
The removal of the external legs of $\, P(r,s)$ is 
nontrivial because of the Mehler propagators and amounts to
\begin{equation}
 \Pi(p,q) =\frac{1}{(2\pi)^4}\int dr \, ds\, G^{-1}(p,-r)P(r,s)G^{-1}(-s,q). \label{PI}
\end{equation}
In fact, it simplifies the final result as it decreases the number 
of the Mehler-kernel factors, that is, the number of parameter integrals.

Due to recurrence relations (\ref{recur})
all field contraction reduce to contractions of the scalar fields.
Let us introduce  shorthand notation for multiple contractions.
In the case of two Mehler kernels we denote
\begin{equation}
 K_2(r,s,p,q)=K(r,s)K(p,q)+K(r,p)K(s,q)+K(r,q)K(s,p) .
\end{equation}
When  there are several external momenta (in this case $r$ and $s$),
 we separate them from the internal ones by a vertical line
 and write
 \begin{equation}
 K_2(r,s|p,q)=K(r,p)K(s,q)+K(r,q)K(s,p).
\end{equation}
With $m$ external and $n$ internal momenta ($n\geq m$),
this generalizes to $K_{m+n}$, defined as
\begin{multline}
2^{\frac{n-m}{2}}\, \bigg(\frac{n-m}{2}\bigg)!\, K_{m+n}(r_1,\,\dots\,,r_m|p_1,\,\dots\,,p_n)=
\\
=\sum_{
\pi_p}
K(r_1,p_{\pi_1})K(r_2,p_{\pi_2}) \dots K(r_m,p_{\pi_m})K(p_{\pi_{m+1}},p_{\pi_{m+2}}) 
\dots K(p_{\pi_{ n-1}},p_{\pi_{n}})
\nonumber
\end{multline}
where $\pi_p$ are permutations of the internal momenta.
$K_{m+n}$ is symmetric under exchange of any two internal 
or any two external momenta.
Calculation of contractions can also be
aided by  recurrence relation 
\begin{multline}
K_{m+n}(r_1,\,\dots\,,r_m|p_1,\,\dots\,,p_n)=
\\
=\sum_{i=1}^n
K(r_1,p_i)K_{m+n-1}(r_2,\,\dots\,,r_m|p_1,\,\dots\,,p_{i-1},p_{i+1},\,\dots\,,p_n).
\end{multline}

 Applying  (\ref{recur}), the sum of 
one-loop contributions can be simplified to
\begin{align}
P_{\phi\phi}&=\sum_{i \le j \le 6} 
(2-\delta_{ij})P_{\phi\phi, ij}+\sum_{7 \le k \le 10}P_{\phi\phi, k}
\\
P^\alpha_{\phi A}&=\sum_{i \le j \le 6} 
(2-\delta_{ij})P^\alpha_{\phi A ,ij}+\sum_{7 \le k \le 10}
P^\alpha_{\phi A, k}
=-i\mu\, \frac{\tilde{r}^\alpha}{r^2}\, P_{\phi\phi}+P'^\alpha_{\phi A}
\\[4pt]
P^{\alpha\beta}_{AA}&=\sum_{i \le j \le 6} 
(2-\delta_{ij})P^{\alpha\beta}_{AA ,ij}+\sum_{7 \le k \le 10}
P^{\alpha\beta}_{AA, k} \nonumber \\
&=-\mu^2\frac{\tilde{r}^\alpha\tilde{s}^\beta}{r^2s^2}\, P_{\phi\phi}
-i\mu\, \frac{\tilde{r}^\alpha}{r^2}\, P'^\beta_{\phi A}
-i\mu\, \frac{\tilde{s}^\beta}{s^2}\, P'^\alpha_{\phi A}
+P'^{\alpha\beta}_{AA},
\end{align}
where due to  similarity of vertices 4 and 5
we have
\begin{align}
&P_{\phi\phi, i5}=-P_{\phi\phi,i4}+P'_{\phi\phi, i5}\label{P'1} \\
&P^{\alpha}_{\phi A,i5}=-P^{\alpha}_{\phi A,i4}-i\mu\frac{\tilde{r}^\alpha}{r^2}
P'_{\phi\phi ,i5}
+P'^\alpha_{\phi A,i5}        \label{P^'2}           \\
&P^{\alpha\beta}_{AA,i5}=-P^{\alpha\beta}_{AA,i4}-\mu^2
\frac{\tilde{r}^\alpha\tilde{s}^\beta}{r^2s^2}P'_{\phi\phi ,i5}
-i\mu\,\frac{\tilde{r}^\alpha}{r^2}P'^\beta_{\phi A, i5}
-i\mu\,\frac{\tilde{s}^\beta}{s^2}P'^\alpha_{\phi A, i5}\label{P'3}
\end{align}
with $i=1,\ldots,5$. This leads to  significant cancellation and 
absorption of terms.

In principle we have two kinds of divergent one-loop contributions to the propagators.
The four-vertices give first order divergences
which were found in \cite{Buric:2012bb}:
\begin{equation}
 \int \phi\phi, \qquad
 \int  A^\mu A_\mu,\qquad
\int \epsilon^{\mu\nu} x_\mu A_\nu\, \phi .
\label{earlier}
\end{equation}
There are also second order contributions from the
 three-vertices which we  calculate here. It is clear that,
 having so many types of interactions, there will be a large number of
 terms. We shall therefore not attempt to present our calculation 
in its full extent,  but  we will rather explain its logic and go through 
the main steps. Some parts of the calculation are straightforward 
albeit long; but to extract and quantify the final results we have
to define a specific prescription adjusted to divergent multiple parameter 
integrals of rational expressions.

Let  us first consider  $P_{\phi\phi}$ which is the
most divergent of the matrix elements. 
The diagram containing two vertices 1 is given by
\begin{align}
P_{\phi\phi, 11}=
&-\frac{4\epsilon^2\mu^2 g^2}{(2\pi)^8}
\int \dif p\,\dif q\,\dif k\,\dif p'\dif q'\dif k'
\delta(p+q+k)\delta(p'+q'+k')
\cos\frac{p\wedge q}{2} \cos\frac{p'\wedge q'}{2}
\nonumber
\\
&\times\epsilon^{\rho\sigma}p_\rho\epsilon^{\mu\nu}p'_\mu\,
\Big\langle
\phi(r) \, \phi(s) \, A_\sigma(p) \, \phi(q) \, \phi(k) \, A_\nu(p') \,
\phi(q') \, \phi(k')
\Big\rangle,
\end{align}
where the correlation function $\, \langle
\phi(r)\phi(s)A_\sigma(p)\phi(q)\phi(k)A_\nu(p')\phi(q')\phi(k')\rangle \, $
is  a sum of contractions of external fields 
with fields in the vertices. There are  90 terms of the type 
$\, \contraction[0.5ex]{}{\phi}{(r)\phi(s)}{A_\sigma}
\bcontraction[0.5ex]{\phi(r)}{\phi}{(s)A_\sigma(p)}{\phi}
\phi(r)\phi(s)A_\sigma(p)\phi(q)
\bcontraction[0.5ex]{}{\phi}{(k)}{A_\nu}
\bcontraction[0.5ex]{\phi(k)A_\nu(p')}{\phi}{(k)}{\phi}
\phi(k)A_\nu(p')\phi(q')\phi(k')\, $
which sum up to 
$\ K_4(r,s|p,q,k,p'\!,q'\!,k')$ and  12  terms with  the
$\, \bcontraction[0.5ex]{}{A}{}{A}AA$ contractions which 
produce $\, K_3(r,s|p,q,p'\!,q')$:  $K_3$ and 
$K_4$  play the role of the usual symmetry factors. We find
\begin{align}
P_{\phi\phi, 11}=&-\frac{4\epsilon^2\mu^2 g^2}{(2\pi)^8}
\int \dif p\,\dif q\,\dif k\,\dif p'\dif q'\dif k'
\delta(p+q+k)\delta(p'\!+q'\!+k')\\
&\times
\cos\frac{p\wedge q}{2} \cos\frac{p'\!\wedge q'}{2}
K_4(r,s|p,q,k,p'\!,q'\!,k') \nonumber	
\\ 
&+\frac{4\epsilon^2 g^2}{(2\pi)^6a}
\int \dif p\,\dif q\,\dif p'\dif q'
\delta(p+q+p'\!+q')
\cos\frac{p\wedge q}{2} \cos\frac{p'\!\wedge q'}{2}
K_3(r,s|p,q,p'\!,q').\nonumber
\end{align}
Contributions of other vertices to the $P_{\phi\phi}$ propagator are
given in Appendix 1.
Representing the $\bcontraction[0.5ex]{}{\phi}{}{\phi}\phi\phi$ 
propagator by a straight line, the $\bcontraction[0.5ex]{}{A}{}{A}AA$ 
by a wiggly line and the $\bcontraction[0.5ex]{}{\phi}{}{A}\phi A$ 
 by a mixed line, these contributions correspond to  diagrams
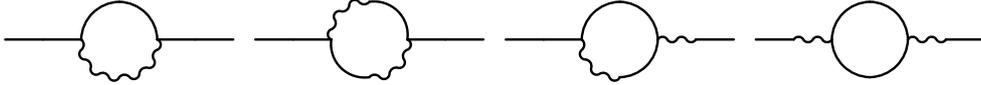
\begin{center}
\begin{fmffile}{propagator1}
 \begin{fmfgraph}(30,20)
 \fmfleft{p,i,m}\fmfright{q,o,n}
 \fmf{phantom,tension=0.5}{p,t}\fmf{phantom,tension=0.5}{t,q}
  \fmf{phantom,tension=0.5}{m,b}\fmf{phantom,tension=0.5}{b,n}
 \fmf{plain,tension=1}{i,v}
 \fmf{plain,tension=1}{v,v1}
 \fmf{wiggly,right=0.5,tension=0.5}{v1,t}
 \fmf{wiggly,right=0.5,tension=0.5}{t,v2}
 \fmf{plain,right=0.4,tension=0.5}{v2,b}
 \fmf{plain,right=0.4,tension=0.5}{b,v1}
 \fmf{plain,tension=1}{v2,v3}
 \fmf{plain,tension=1}{v3,o}
 \end{fmfgraph}
\begin{fmfgraph}(30,20)
 \fmfleft{p,i,m}\fmfright{q,o,n}
 \fmf{phantom,tension=0.5}{p,t}\fmf{phantom,tension=0.5}{t,q}
  \fmf{phantom,tension=0.5}{m,b}\fmf{phantom,tension=0.5}{b,n}
 \fmf{plain,tension=1}{i,v}
 \fmf{plain,tension=1}{v,v1}
 \fmf{plain,right=0.5,tension=0.5}{v1,t}
 \fmf{wiggly,right=0.5,tension=0.5}{t,v2}
 \fmf{plain,right=0.5,tension=0.5}{v2,b}
 \fmf{wiggly,right=0.5,tension=0.5}{b,v1}
 \fmf{plain,tension=1}{v2,v3}
 \fmf{plain,tension=1}{v3,o}
 \end{fmfgraph}
\begin{fmfgraph}(30,20)
 \fmfleft{p,i,m}\fmfright{q,o,n}
 \fmf{phantom,tension=0.5}{p,t}\fmf{phantom,tension=0.5}{t,q}
  \fmf{phantom,tension=0.5}{m,b}\fmf{phantom,tension=0.5}{b,n}
 \fmf{plain,tension=1}{i,v}
 \fmf{plain,tension=1}{v,v1}
 \fmf{wiggly,right=0.5,tension=0.5}{v1,t}
 \fmf{plain,right=0.4,tension=0.5}{t,v2}
 \fmf{plain,right=0.4,tension=0.5}{v2,b}
 \fmf{plain,right=0.4,tension=0.5}{b,v1}
 \fmf{wiggly,tension=1}{v2,v3}
 \fmf{plain,tension=1}{v3,o}
 \end{fmfgraph}
  \begin{fmfgraph}(30,20)
 \fmfleft{p,i,m}\fmfright{q,o,n}
 \fmf{phantom,tension=0.5}{p,t}\fmf{phantom,tension=0.5}{t,q}
  \fmf{phantom,tension=0.5}{m,b}\fmf{phantom,tension=0.5}{b,n}
 \fmf{plain,tension=1}{i,v}
 \fmf{wiggly,tension=1}{v,v1}
 \fmf{plain,right=0.4,tension=0.5}{v1,t}
 \fmf{plain,right=0.4,tension=0.5}{t,v2}
 \fmf{plain,right=0.4,tension=0.5}{v2,b}
 \fmf{plain,right=0.4,tension=0.5}{b,v1}
 \fmf{wiggly,tension=1}{v2,v3}
 \fmf{plain,tension=1}{v3,o}
 \end{fmfgraph}
 \end{fmffile}
\end{center}
The one-loop quantum correction is the sum of all enumerated terms. 
In order to calculate it one first expresses Mehler kernels $K_3$, 
$K_4$ etc. as parameter integrals which are Gaussian in $p$, $q$, $k$; 
momentum integration can then be performed\footnote{This is 
true if the denominator of the rational expression which appears as a
factor in the course of integrations is of relatively low degree: 
otherwise one has to introduce additional Schwinger parametrizations.}. The result 
is a divergent multiple parameter integral which, except in the simplest 
cases, cannot be done explicitly when there is more than one integration. 

A more useful way to analyze divergences is to use the
amputated propagators, in which $K_3$ and $K_4$ for example reduce to
$K$ and $K_2$. But in order to find the amputated 
propagators we  need to find all matrix elements, that is $P_{\phi A}$ and 
$P_{AA}$ as well. They are of a form similar to  $P_{\phi \phi}$ and 
somewhat longer; the full expressions are given in  \cite{LukaPhD}.
Multiplication of the amputated propagator by a multiplet 
of classical external fields gives the one-loop effective action:
\begin{equation}
  \Gamma=\frac{1}{2}\int \mathrm{d} r \, \mathrm{d} s \, 
\Phi^T(-r)\, {\Pi}(r,s)\,\Phi(-s) .    \label{EFF}
 \end{equation}
 
Our main goal is to extract divergent parts of the
last expression. We find
\begin{align}
&\Pi^{\mu\nu}(p,q)=a^2 p^2 q^2 P'^{\mu\nu}_{AA}(p,q)      \label{+}
\\[8pt] 
&\Pi^\mu(p,q)=-ia^2\mu p^2 \tilde{q}_\rho P'^{\mu\rho}_{AA}(p,q)
-\frac{a p^2}{(2\pi)^2} \int \dif k\, P'^\mu_{\phi A}(p,k) K^{-1}(-k,q)
\\[8pt] 
&\Pi(p,q)=
-a^2 \mu^2 \tilde{p}_\rho \tilde{q}_\sigma P'^{\rho\sigma}_{AA}(p,q)  \label{++}
\\[6pt]
&\qquad\quad+\frac{ia\mu  \tilde{p}_\rho}{(2\pi)^2} 
 \int \dif k\, P'^\rho_{\phi A}(p,k)
K^{-1}(-k,q) 
+\frac{ia\mu \tilde{q}_\rho}{(2\pi)^2}  \int \dif k\,
 P'^\rho_{\phi A}(q,k)
K^{-1}(-k,p) 
\nonumber\\[6pt]
&\qquad\quad +\frac{1}{(2\pi)^4} \int \dif p' \dif q'\,
K^{-1}(p,-p')P_{\phi\phi}(p',q')K^{-1}(-q',q)  . \nonumber
\end{align}
In comparison to full propagators, these
 expressions are  considerably simpler. For  exact forms of
 $\, P'^\rho_{\phi A}$ and $\, P'^{\rho\sigma}_{AA}$ we
 refer to \cite{preparation}.
 
\initiate

\section{Divergences}

\subsection{The $\phi\phi$-sector}

Part of the  effective action which gives the one-loop quantum 
corrections to the propagators is given by (\ref{EFF}).
In the usual case, 2-point functions  have the form
\begin{equation}
\Pi(r,s) = \Pi(r) \, \delta(r+s)   \label{delta}
\end{equation}
which reflects the translational invariance. (In our convention
for the Fourier transformation, all momenta are incoming.)
However, we are dealing with a nonlocal action which is
not translationally invariant. Therefore
in order to  recover the form of divergences in the effective
action in position space we  introduce the so-called `short' and `long 
variable', respectively $u$ and $v$:
\begin{equation}
 u=\frac{r+s}{2}, \qquad v=\frac{r-s}{2} .
\end{equation}
Here $u$ denotes the difference between the incoming and 
outgoing momenta in a vertex or along a line.  In  translationally 
invariant case one integrates over $u$ and 
the  divergences remain in $ \Pi(v)$,
\begin{equation}
 \Gamma
= \int \mathrm{d} u \, \mathrm{d} v \, 
\Phi^T(-u-v)\, {\Pi}(u+v)\delta(2u)\,\Phi(-u+v)=
\frac{1}{2}\int \mathrm{d} v \, 
\Phi^T(-v)\, {\Pi}(v)\,\Phi(v). \nonumber
\end{equation}
Here the  $\delta$-fuction  is smeared,  roughly replaced 
by an exponentially decreasing factor
\begin{equation}
 \delta(u) = \lim_{\sigma\to 0} \frac{1}{2\pi \sigma^2}\,  \label{dddel}
 e^{-\frac{u^2}{2\sigma^2}} ,
\end{equation}
which is hidden in  parameter integrations. The exponential 
factors regularize  all momentum integrations in the UV sector:
divergences occur   in the IR sector, for small values of $u$. Our
 strategy to calculate them is as follows. We expand terms in the 
 effective action (\ref{EFF}) around $u=0$, keeping all parameter 
 integrals which come from the Mehler kernels and  Schwinger 
 parametrizations. This gives momentum integrals of the Poisson type 
 (which one can calculate) and usually leaves  two parameter integrations.  
 In order to identify the types of divergences we introduce appropriate 
 regulators, expand fields in powers of $u_\alpha$  as in (\ref{__}),
and integrate term by term: only the first few terms are infinite.
As mentioned, we consider only the lower bound in momentum integrals as 
that is where divergences lie. Eventually,
 we find new nonlocal divergences of the form
\begin{equation}
 \int  \phi\, \Box^{-1} \phi, \qquad
\int  \phi\, \Box^{-2} \phi.   \label{Box-2}
\end{equation}

Let us discuss  details of the calculation of $\, \Gamma_{\phi\phi}^{(div)}$.
After removal of the external legs 
we obtain  lengthy expression which contains several 
hundred terms. Most of them are finite, which
can be checked by  power counting. We consider divergent terms
 in the increasing order of powers of the momentum, expecting
 to find in the lowest order only mass and wave function renormalizations.
However, a closer inspection shows that divergent terms of the 
lowest degree can, and do, contain nonlocal $\, \Box^{-2}$ and $\Box^{-1}$ terms
which are new. We focus therefore on them; we
 denote the corresponding parts of $\,\Pi$, $\Gamma$ by a tilde.
Parts of the amputated propagator  $\Pi$ which contain nonlocal
divergent contributions are:
\begin{align}
\tilde\Pi^{(1)}_{\phi\phi}&=\!-\, \frac{32 a\mu^8g^2}{(2\pi)^2\epsilon^2}\,
 \frac{r\wedge s}{r^2s^2(r+s)^2}\, \sin\frac{r\wedge s}{2}\!\!
\int \!\! dp\,dq\,\delta(-r-s+p+q) 
\sin\frac{p\wedge q}{2}\, 
\frac{p\wedge q}{p^2} \, K(p,q) \nonumber
\\[4pt]
\tilde\Pi^{(2)}_{\phi\phi}&=\!-\, \frac{8 a\mu^8g^2}{(2\pi)^2\epsilon^2}\, 
 \frac{1}{r^2s^2}
\int \!\! dp\,dq\,
 \delta(-r-s+p+q)
\sin\frac{p\wedge r}{2}\sin\frac{q\wedge s}{2}
\, \frac{(p\wedge r)\,(q\wedge s)}{p^2q^2} \,K(p,q)\nonumber\\[4pt]
  \tilde\Pi^{(3)}_{\phi\phi}&=\frac{8 \mu^4g^2}{(2\pi)^2}\frac{1}{r^2s^2}
\int \!\! dp\,dq\,
 \delta(-r-s+p+q)
 \sin\frac{p\wedge r}{2}\sin\frac{q\wedge s}{2}\,
 \frac{(p\cdot r)(q\cdot s)}{p^2q^2}\, K(p,q)\nonumber\\[4pt]
  \tilde\Pi^{(4)}_{\phi\phi}&=\!-\, \frac{8 a\mu^8g^2}{(2\pi)^2\epsilon^2}\, 
  \frac{1}{r^2s^2}
\int \!\! dp\,dq\,
 \delta(-r-s+p+q)
\sin\frac{p\wedge r}{2}\sin\frac{q\wedge s}{2}\,
\frac{(p\wedge r)\,(q\wedge s)}{p^2(p-r)^2} \, K(p,q)\nonumber
\\[2pt] &\ \ +(r \leftrightarrow s)\nonumber  .
\end{align}
Introducing the short and long variables and 
expressing the Mehler kernel in  parametric form we find
the following contributions to the effective action:
\begin{align}
 \tilde\Gamma_{\phi\phi}^{(1)}&=\frac{2ag^2}{\pi} \,\Re\!
 \int du\,dv\, \frac{\phi(-u-v)\phi(-u+v)}{(u+v)^2(u-v)^2 u^2} \,
 (v\cdot\tilde{u})\tilde{u}_\alpha \, e^{-iu\wedge v}
 \int\limits^{\infty}_{1}\frac{d\xi}{\xi}\frac{\xi-1}{\xi+1} 
 e^{-(\xi+\frac{1}{\xi})\frac{u^2}{2\mu^2}}
 \nonumber \\
 &\times \int dp\, p^\alpha
 \Big(e^{-i\epsilon\frac{p\cdot\tilde{u}}{\mu^2}}-
 e^{i\epsilon\frac{p\cdot\tilde{u}}{\mu^2}}\Big) 
 e^{-\frac{1}{\xi}\frac{p^2}{2\mu^2}+\frac{1}{\xi}\frac{p\cdot u}{\mu^2}}
\nonumber\\[8pt]
 \tilde\Gamma_{\phi\phi}^{(2)}&=\frac{ag^2}{2\pi\mu^2} \,\Re\!
 \int du\,dv\, \frac{\phi(-u-v)\phi(-u+v)}{(u+v)^2(u-v)^2} \, e^{iu\wedge v}
 \int\limits_{1}^{\infty}\frac{d\xi}{\xi}\frac{\xi-1}{\xi+1}
 \int\limits_{0}^{\infty}d\eta\, 
 e^{-(\xi+\frac{1}{\xi}+4\eta)\frac{u^2}{2\mu^2}}
 \nonumber \\
 &\times \int dp \, 
 \frac{(p\cdot(\tilde{u}+\tilde{v}))(p\cdot(\tilde{u}-\tilde{v})
 +2u\cdot\tilde{v})}{p^2} \,
  e^{-(\frac{1}{\xi}+\eta)\frac{p^2}{2\mu^2}+(\frac{1}{\xi}+2\eta)
  \frac{p\cdot u }{\mu^2}}
 \Big(e^{-i\epsilon\frac{p\cdot\tilde{v}}{\mu^2}}
 -e^{i\epsilon\frac{p\cdot\tilde{u}}{\mu^2}}\Big)
\nonumber \\[8pt]
\tilde\Gamma_{\phi\phi}^{(3)}&=\frac{g^2}{2\pi\mu^2}\,\Re\!
 \int du\,dv\, \frac{\phi(-u-v)\phi(-u+v)}{(u+v)^2(u-v)^2} \, e^{iu\wedge v}
 \int\limits_{1}^{\infty}\frac{d\xi}{\xi}\frac{\xi-1}{\xi+1}
 \int\limits_{0}^{\infty}d\eta\, 
 e^{-(\xi+\frac{1}{\xi}+4\eta)\frac{u^2}{2\mu^2}}
  \nonumber \\
 &\times \int dp \, 
 \frac{(p\cdot(u+v))((2u-p)\cdot(u-v))}{p^2} \,
  e^{-(\frac{1}{\xi}+\eta)\frac{p^2}{2\mu^2}+(\frac{1}{\xi}+2\eta)
  \frac{p\cdot u }{\mu^2}}
 \Big(e^{i\epsilon\frac{p\cdot\tilde{u}}{\mu^2}}
 -e^{-i\epsilon\frac{p\cdot\tilde{v}}{\mu^2}}\Big)
\nonumber\\[8pt]
 \tilde\Gamma_{\phi\phi}^{(4)}&=\frac{ag^2}{2\pi\mu^2} \,\Re\!
 \int du\,dv\, \frac{\phi(-u-v)\phi(-u+v)}{(u+v)^2(u-v)^2} \, e^{iu\wedge v}
 \int\limits_{1}^{\infty}\frac{d\xi}{\xi}\frac{\xi-1}{\xi+1}
 \int\limits_{0}^{\infty}d\eta\, 
e^{-(\xi+\frac{1}{\xi}+\eta)\frac{u^2}{2\mu^2}-\eta\frac{v^2+2u\cdot v}{2\mu^2}}
 \nonumber \\
&\times \int dp \, 
 \frac{(p\cdot(\tilde{u}+\tilde{v}))(p\cdot(\tilde{u}-\tilde{v})
 +2u\cdot\tilde{v})}{p^2} \,
  e^{-(\frac{1}{\xi}+\eta)\frac{p^2}{2\mu^2}+(\frac{1}{\xi}+\eta)
  \frac{p\cdot u}{\mu^2}+\eta\frac{p\cdot v}{\mu^2}}
 \Big(e^{-i\epsilon\frac{p\cdot\tilde{v}}{\mu^2}}
 -e^{i\epsilon\frac{p\cdot\tilde{u}}{\mu^2}}\Big). \nonumber
\end{align}

In order to analyze the behavior of these integrals, we first perform
 the Gaussian integration over  $p$. 
For $ \tilde\Gamma_{\phi\phi}^{(1)} $, which is the simplest, we obtain
\begin{align}
 \Gamma_{\phi\phi}^{(1)}&=-2a\epsilon g^2\int du\,dv\, 
 \frac{\phi(-u-v)\phi(-u+v)}{(u+v)^2(u-v)^2 u^2}
 (u\cdot\tilde{v})\sin(u\wedge v)
 \int\limits^{\infty}_{1}d\xi\,\frac{\xi-1}{\xi+1} \, 
 e^{-(1+\epsilon^2)\xi\frac{u^2}{2\mu^2}}. \nonumber
\end{align}
We need to estimate this expression at the lower
bound $u=0$, so we expand field $\phi$ around this point,
\begin{equation}
\phi(-u+v) = \phi(v)-\p_\alpha\phi(v) \, u^\alpha +\dots
\label{__}
\end{equation}
The leading order term is
\begin{align}
 \tilde\Gamma_{\phi\phi}^{(1)}&=-\frac{2a\epsilon^2 g^2}{\mu^2}\int dv\, 
\frac{\phi(-v)\phi(v)}{v^2}\int du
 \int\limits^{\infty}_{1}d\xi\,\frac{\xi-1}{\xi+1} \, 
 e^{-(1+\epsilon^2)\,\xi\frac{u^2}{2\mu^2}}.
 \label{454}
\end{align}
One can easily see that this expression is divergent, that is, that the
 result of the last two integrations at the lower $u$-bound is infinite:
we either  put  $u=0$, in which case the
$\xi$-integral,  $\ \int^{\infty}_{1}d\xi\,(\xi-1)/(\xi+1)$,
 is divergent at  $\xi=\infty$, or we first perform the 
 $\xi$-integration and obtain $\, \int du\, e^{-u^2}/u^2$,
 which is  logarithmically divergent at $u=0$. 
 
 Using the
regularization described  in Appendix 4, for the divergent part 
of $\tilde\Gamma^{(1)}_{\phi\phi}$ we obtain
\begin{equation}
\tilde\Gamma_{\phi\phi}^{(1,div)}=
-\frac{16\pi^3a\epsilon^2g^2}{1+\epsilon^2}\log\Lambda\int \phi\,\Box^{-1}\phi,
\label{Gamma1}
\end{equation}
where $\Lambda $ is the regularization parameter.
The analysis of the remaining three terms is similar albeit 
more complicated, as the corresponding leading-order expressions,
after expansion in $u$, contain integration in two parameters
$\xi$ and $\eta$; relevant terms are written in Appendix 3. 
 A systematic procedure which enables to estimate these 
 integrals, that is to introduce a regulator and sum up different 
 contributions,  is described in Appendix 4.
Adding all  divergent nonlocal contributions in the $\phi\phi$-sector  we obtain
\begin{align}
\tilde\Gamma_{\phi\phi}^{(div)}=
\left(\frac{8}{\epsilon^2}-14+\epsilon^2\right)\pi^3\mu^4 g^2\log\Lambda
\int  \phi\,\Box^{-2}\phi
 +\epsilon^2\pi^3\mu^2 g^2\Lambda^2
\int \phi\,\Box^{-1}\phi .
\end{align}
In addition,    $ \Gamma_{\phi\phi}^{(div)}$ contains the
$\,\, \displaystyle{\int \phi\phi} \, $ term found 
before with a corrected infinite prefactor.

\subsection{The $AA$-sector}

The most important obstacle in constructing renormalizable 
noncommutative gauge theory on the Moyal space 
is quadratically divergent IR term of the form 
$\Pi^{\mu\nu}\propto p^\mu p^\nu/(p^2)^2$  which
comes from the non-planar part of the gauge-field self energy 
\cite{Hayakawa:1999zf,Ruiz:2000hu,Attems:2005jt}
and seems to be independent on the gauge fixing.
It gives rise to a nonlocal counterterm,
\cite{Blaschke:2008yj,Blaschke:2009gm}
\begin{equation}
 \int  F^{\mu\nu}\star\frac{1}{D^2 \tilde{D}^2}\star F_{\mu\nu}.
\end{equation}
 As we will see, there is no such term in our theory, but 
other nonlocal terms exist.

Analyzing the form of $\Pi_{\mu\nu}$
we find only two amputated-propagator 
terms in  the $AA$ sector which can be sources of nonlocal divergences:
\begin{align}
&
\tilde\Pi_{\mu\nu}^{(1)}(r,s)=\frac{4a\mu^2 g^2}{(2\pi)^2}
 \int dp\,dq\,\delta(-r-s+p+q)\sin\frac{p\wedge r}{2}
 \sin\frac{q\wedge s}{2} \, \frac{p_\mu q_\nu}{p^2 q^2} \, K(p,q) \nonumber
\\
&
 \tilde\Pi_{\mu\nu}^{(2)}(r,s)=-\frac{8a\mu^2 g^2}{(2\pi)^2}
 \int dp\,dq\,\delta(-r-s+p+q)\sin\frac{p\wedge r}{2}
 \sin\frac{q\wedge s}{2}\, \frac{\tilde{p}_\mu \tilde{q}_\nu}{p^2 q^2}
 \, K(p,q)  . \nonumber
\end{align}
In fact they are, up to replacement $p_\mu\to \tilde p_\mu$,
$q_\nu\to \tilde q_\nu$, almost the same and they have the same divergent 
parts: we therefore analyze only the first.
The  computational details are very similar to those
which we developed and explained in Appendix 4 for the $\phi\phi$-sector.
As before, we want to examine the behavior of the integrals for small $u$. 
Introducing the short and  long variables, $\,\tilde\Pi_{\mu\nu}^{(1)}(r,s) $
becomes
\begin{equation*}
 \tilde\Pi_{\mu\nu}^{(1)}=\frac{a\mu^2g^2}{\pi^2} \int dp\,\sin\frac{p\wedge (u+v)}{2}
\, \sin\frac{(2u-p)\wedge(u-v)}{2}\frac{p_\mu(2u-p)_\nu}{p^2(2u-p)^2}\, K(p,2u-p) .
\end{equation*}
Using the Schwinger parametrization and expressing the Mehler kernel in the
parameter form, we obtain
\begin{align}
\tilde\Pi_{\mu\nu}^{(1)}(u,v)= -\frac{a}{8\pi\mu^4} 
 &\int dp\,\big(\cos(p\wedge u+u\wedge v)-\cos(p\wedge v-u\wedge v)\big) \,
 \frac{2p_\mu u_\nu-p_\mu p_\nu}{p^2} 
 \nonumber
 \\
 \times
 &\int\limits^{\infty}_{0}d\eta\, e^{-\eta\frac{(2u-p)^2}{2\mu^2}}
 \int\limits^{\infty}_{1}\frac{d\xi}{\xi}\,\frac{\xi-1}{\xi+1} \, 
 e^{-\frac{1}{2\mu^2}\big(\xi u^2+\frac{1}{\xi}(p-u)^2\big)}.
\end{align}
As we wish to single out  terms proportional to 
$\tilde{v}^\mu\tilde{v}^\nu/(v^2)^2$
we can neglect the first cosine. After the Gaussian integration
we obtain 
\begin{align}
 \tilde\Gamma_{AA}^{(1)}=& -\frac{ag^2}{8\mu^2} \, \Re \! 
 \int du\,dv\, A_\mu(-u-v)A_\nu(-u+v) \, e^{iu\wedge v} \!\! 
 \int\limits^{\infty}_{1} d\xi \,\frac{\xi-1}{\xi+1} \, 
 e^{-\xi\frac{u^2}{2\mu^2}}
  \nonumber\\
  &\times 
 \int\limits^{\infty}_{0}d\eta\,
 e^{-\frac{\eta}{1+\eta\xi}\frac{u^2}{2\mu^2}}
 e^{-\frac{\xi\epsilon^2}{1+\eta\xi}\frac{v^2}{2\mu^2}+
 i\epsilon\frac{1+2\eta\xi}{1+\eta\xi}
 \frac{u\cdot\tilde{v}}{2\mu^2}}
 \nonumber\\
  &\times 
 \Bigg(
 \frac{(1+2\eta\xi)^2u^\mu u^\nu+2i\epsilon\xi\tilde{v}^\mu u^\nu}
 {(1+2\eta\xi)^2u^2-\xi^2\epsilon^2 v^2+i\epsilon\xi(1+2\eta\xi)(u\cdot\tilde{v})}\, +
 \nonumber\\
  &\qquad
 +\frac{\xi\mu^2}{(1+2\eta\xi)^2u^2-\xi^2\epsilon^2 v^2
 +i\epsilon\xi(1+2\eta\xi)(u\cdot\tilde{v})}\,
 \times
 \nonumber
 \\[8pt]
 &\qquad \ 
 \times\Big(\delta^{\mu\nu}
+ \frac{2(1+2\eta\xi)^2u^\mu u^\nu-2\epsilon^2\xi^2\tilde{v}^\mu\tilde{v}^\nu
+i\epsilon\xi(1+2\eta\xi)(u^\mu\tilde{v}^\nu
 +u^\nu\tilde{v}^\mu)}
 {(1+2\eta\xi)^2u^2-\xi^2\epsilon^2 v^2+i\epsilon\xi(1+2\eta\xi)(u\cdot\tilde{v})}
 \nonumber
 \\[8pt]
 &\qquad \quad
 +\frac{2(1+2\eta\xi)^2 u^\mu u^\nu-2\epsilon^2\xi^2\tilde{v}^\mu\tilde{v}^\nu
 +i\epsilon\xi(1+2\eta\xi)(u^\mu\tilde{v}^\nu
 +u^\nu\tilde{v}^\mu)}{2\xi(1+\eta\xi)\mu^2}
 \Big)
 \Bigg) .
 \nonumber
\end{align}
The singular part of this long expression is 
in fact quite simple,
\begin{equation}
 \tilde\Gamma_{AA}^{(1,div)} =
 \frac{ag^2}{8\epsilon^2}
 \int du\,dv\, \frac{A_\mu(-v)A_\nu(v)}{v^2}
 \int\limits^{\infty}_{1} \frac{d\xi}{\xi} \,
 \frac{\xi-1}{\xi+1} \, e^{-\xi\frac{u^2}{2\mu^2}}
 \int\limits^{\infty}_{0} d\eta\, 
 \left(\delta_{\mu\nu}+2\frac{\tilde{v}_\mu\tilde{v}_\nu}{v^2}
 -\frac{\epsilon^2\xi}{1+\eta\xi}\frac{\tilde{v}_\mu\tilde{v}_\nu}{\mu^2}\right) ,
 \nonumber
\end{equation}
so in the $\Lambda$-leading order  we obtain
\begin{align}
 \Gamma_{AA}^{(1,div)}=
 \frac{ag^2}{8\epsilon^2}
 \int du\,dv\, \frac{A^\mu(-v)A_\mu(v)}{v^2}
 \int\limits^{\infty}_{1} \frac{d\xi}{\xi} \, e^{-\xi\frac{u^2}{2\mu^2}}
 \int\limits^{\infty}_{0} d\eta\, 
 =\frac{a\pi^3\mu^2 g^2}{\epsilon^2}\,\beta\Lambda\log\Lambda
 \int A^\mu\,\Box^{-1}\!A_\mu.       \label{460}
\end{align}
Adding $\, \Gamma_{AA}^{(2,div)}$ to (\ref{460}) we find
\begin{equation}
 \Gamma_{AA}^{(div)}=-\frac{a\pi^3\mu^2 g^2}{\epsilon^2}\,\beta\Lambda\log\Lambda
 \int dx\, A^\mu(x)\,\Box^{-1}\!A_\mu(x),
\end{equation}
which after setting $\beta=1$ becomes
\begin{equation}
 \Gamma_{AA}^{(div)}=-\frac{a\pi^3\mu^2g^2}{\epsilon^2}\, \Lambda\log\Lambda
 \int  A^\mu\,\Box^{-1}\!A_\mu.
\end{equation}

\initiate

\section{Conclusion and outlook}

We calculated in this paper the one-loop corrections 
to the propagators in a dimensionally reduced Yang-Mills gauge 
theory defined on the truncated Heisenberg space. 
The classical action  is given by (\ref{Kin}-\ref{Sint}) and
 the theory is perturbatively quantized around its vacuum
solution $\phi =0$, $A_\mu =0$.
In the previous paper \cite{Buric:2012bb} we found the one-loop divergences 
of the effective action of the first order. They comprise  tadpoles
\begin{equation}
\int \phi, \qquad \int \epsilon^{\mu\nu} x_\mu A_\nu,  \label{TAD}
\end{equation}
and  mass terms
\begin{equation}
\int \phi \phi,\qquad \int A_\mu A^\mu,\qquad \int  
\epsilon^{\mu\nu} x_\mu A_\nu \phi.                       \label{MAS}
\end{equation}
Here we calculated  the one-loop
divergent corrections of the second order to the $\phi\phi$ and
$AA$ propagators and found the following additional  terms:
\begin{equation}
\int \phi\,\Box^{-2}\phi, \qquad \int \phi\,\Box^{-1}\phi, \qquad 
\int A_\mu\Box^{-1}A^\mu.                            \label{NON}
\end{equation}
We have not calculated the $\phi A$ one-loop divergences, but 
from symmetry we  expect that there are nonvanishing 
 nonlocal corrections in this sector too.

The result is not what we expected or hoped for. Namely, in  related 
models with scalar and spinor matter
it was possible to attribute renormalizability to the
background geometry, that is, to an adequate inclusion 
of  geometric quantities in the  Lagrangian,
\cite{Buric:2015vja}. It is known on the other hand that
 on commutative curved spaces scalar and spinor theories are renormalizable 
only if matter is nonminimally coupled to the background 
curvature and torsion, \cite{Shapiro:2001rz}, and
this pattern is exactly followed in the Grosse-Wulkenhaar and 
Vignes Tourneret models. We expected a similar behavior of our
noncommutative $U(1)$ model; however, the outcome of our calculation 
proves differently.
 
Gauge theories on noncommutative spaces have
an additional freedom  which 
comes with the existence of  covariant coordinates.
This means that one can include the gauge potentials via
 $\tX$ in the action directly, for example as 
 $\,(\tX_\mu \tX^\mu)^n$ or $\, \exp (\alpha_\mu \tX^\mu) $,
to obtain new classes of theories. Even if one restricts oneself to theories 
written geometrically, that is, by considering
only terms which are proportional to the  volume form, there are new 
gauge-invariant quantities. In our three-dimensional case they are,  
\begin{equation}
\tr \tX (*\tX),\quad
 \tr \tX^3,\quad  
\tr \tX F, \quad  \tr \tX^2(*F) .  \label{traces}
\end{equation}
However, not all of these expressions are independent
because of relation (\ref{FX}), which on the truncated
Heisenberg space reads
\begin{equation}
\tX^2=\tF+\mu( *\tX)-\frac{i\mu^2}{4\epsilon}\, [\theta^1,\theta^2] .
\end {equation}
Calculating  the first two terms of (\ref{traces}) we obtain
\begin{align}
& \tr \tX (*\tX) =\tr\Big(\tX_1^2+\tX_2^2+(1-\epsilon^2)\tX_3^2\Big) \\
& \phantom{\tr \tX (*\tX)} =\tr \Bigg( \frac{(1-\epsilon^2)\mu g}{\epsilon }\, 
\phi +\frac{2\mu^2g}{\epsilon}\,
\epsilon^{\mu\nu} x_\mu A_\nu-(1-\epsilon^2)g^2\phi^2 -g^2 A_\mu A^\mu \Bigg) 
\nonumber\\[8pt]
&
\tr \tX^3=\tr  \Big((3-\epsilon^2)[\tX_1,\tX_2]\tX_3+2i\epsilon 
\tX_3(\tX_1^2+\tX_2^2)\Big) \\[4pt]
& \phantom{\tr \tX^3}= \tr  \Bigg( \frac{(3-\epsilon^2)\mu^2 g}{\epsilon}\, \phi 
+\frac{2\mu^4g}{\epsilon}\,x_\mu x^\mu \phi
+\frac{2\mu^3 g}{\epsilon}\,\epsilon^{\mu\nu}x_\mu A_\nu \nonumber\\[0pt]
&\phantom{\tr \tX^3}\qquad -(3-\epsilon^2)g\phi F_{12}-2\mu^2g^2\epsilon^{\mu\nu}
(x_\mu A_\nu +A_\nu x_\mu)\phi  
-\mu g^2A_\mu A^\mu  \nonumber \\[4pt]
&\phantom{\tr \tX^3}\qquad+2\epsilon g^3 A_\mu A^\mu \phi \Bigg) , \nonumber
\end{align}
where we neglected the  boundary and cosmological terms.
The second pair gives
\begin{align}
&
\tr \tX\tF= \tr \tX^3 -\mu\, \tr \tX (*\tX) -\tr 
\frac{(1-\epsilon^2)\mu^2 g}{2\epsilon}\, \phi , \\[4pt]
& \tr \tX^2(*\tF) =  \tr \tF(*\tF)+\mu\, \tr \tX \tF -\tr 
\frac{(1-\epsilon^2)\mu^3 g}{2\epsilon}\, \phi .
\end{align}
We see therefore that, were only 
divergences (\ref{TAD}-\ref{MAS}) present,
the theory would have been renormalizable as we could expand the 
initial action by adding purely geometric terms.  It 
is also interesting to note that  addition of the new  terms
can translate the classical vacuum $\phi =0$, $A_\mu=0$ 
arbitrarily, which is a point that needs further understanding. 
But obviously, it is not possible to cancel nonlocal divergences (\ref{NON}) 
in this way, using only polynomial expressions of covariant coordinates.
We come again, in this model, across an occurrence of the UV/IR mixing. 
We therefore conclude that geometric gauge 
theories cannot render a renormalizable theory. 

Perhaps a correct way to find a renormalizable gauge model is
to consider nonpolynomial interactions, or to add 
nonlocal terms imposing, as a version of LS duality, symmetry
under exchange $\, \Box\leftrightarrow \Box^{-1}$. 
The latter was implemented for the scalar field theory in \cite{Gurau:2009},
but it was not successful for the gauge fields 
\cite{Blaschke:2008yj,Blaschke:2009gm}. The other direction of research
would be to analyze  a matrix model which corresponds to our
theory. This numerical study could give  important information about 
 properties of the gauge fields, and 
 would enlarge our understanding of the matrix regularizations.

Finally, a possible explanation of nonrenormalizability of  gauge 
theories is that on noncommutative spaces they are intimately related
to gravity. Not only is this seen in the fact that we can combine
momenta $p_\alpha\in\cal{A}$  with the gauge potentials into a unique covariant object. 
The gauge and coordinate transformations in noncommutative case  
cannot be clearly separated: indeed,  infinitesimal local translations 
\begin{equation}
\delta\phi = a^\alpha [p_\alpha, \phi]          \label{t}
\end{equation}
have the same form as infinitesimal $U(1)$ transformations
\begin{equation}
\delta\phi = \epsilon^\alpha [A_\alpha, \phi] ,     \label{u}
\end{equation}
that is, (\ref{t}) is a special case of (\ref{u}).
If  gauge theories are a part of gravity or vice versa, then 
the correct way to understand their renormalizability would be
 to understand the noncommutative gravity first.


\vskip1cm

{\bf Acknowledgement}\  \ 
This work was supported by the Serbian Ministry of Education, Science
and Technological Development Grant ON171031.


\newpage

\noindent
\begin{large}
{\bf Appendix 1}
\end{large}
\vskip0.5cm
\noindent
The rest of the contributions--and the respective diagrams--needed for finding the second order propagator correction $P_{\phi\phi}$ via (\ref{P'1}) are given below.

\begin{align}
P_{\phi\phi ,12}=&-\frac{8\mu^4g^2}{(2\pi)^8}
\int \dif p\,\dif q\,\dif k\,\dif p'\dif q'\dif k'
\delta(p+q+k)\delta(p'\!+q'\!+k') \label{Pff11}\\
&\times\frac{p\wedge q}{2}\, \sin\frac{p\wedge q}{2}\,
 \cos\frac{p'\!\wedge q'}{2} \frac{1}{(p+q)^2}\, 
K_4(r,s|p,q,k,p'\!,q'\!,k')\nonumber
\\ 
&+\frac{8\mu^2g^2}{(2\pi)^6a}
\int \dif p\,\dif q\,\dif p'\dif q'
\delta(p+q+p'\!+q')\nonumber\\
&\times\frac{p\wedge q}{2} \, \sin\frac{p\wedge q}{2} \,
\cos\frac{p'\!\wedge q'}{2}\frac{1}{(p+q)^2}\, 
K_3(r,s|p,q,p'\!,q')\nonumber\\[8pt]
P_{\phi\phi ,13}=&
\frac{8\epsilon^2\mu^4g^2}{(2\pi)^8}
\int \dif p\,\dif q\,\dif k\,\dif p'\dif q'\dif k'
\delta(p+q+k)\delta(p'\!+q'\!+k')\\
&\times\cos\frac{p\wedge q}{2}\, \cos\frac{p'\,\wedge q'}{2}
\frac{(p+q)^\mu}{(p+q)^2}\,
\frac{\partial}{\partial p^\mu}K_4(r,s|p,q,k,p'\!,q'\!,k') 	
\nonumber\\ 
&-\frac{8\epsilon^2\mu^2g^2}{(2\pi)^6a}
\int \dif p\,\dif q\,\dif p'\dif q'
\delta(p+q+p'\!+q')
\cos\frac{p\wedge q}{2}\nonumber\\
&\times\cos\frac{p'\!\wedge q'}{2}\,\frac{(p+q)^\mu}{(p+q)^2}
\frac{\partial}{\partial p^\mu}
K_3(r,s;p,q,p'\!,q') \nonumber\\[8pt]
P'_{\phi\phi, 15}=&\frac{4\mu^4g^2}{(2\pi)^6}
\int \dif p\,\dif q\,\dif p'\dif q'
\delta(p+q+p'\!+q')\\
&\times \frac{p\wedge q}{2}\, \sin\frac{p\wedge q}{2} 
\cos\frac{p'\!\wedge q'}{2}\frac{1}{p^2q^2}\,K_3(r,s|p,q,p'\!,q'),\nonumber
\\[8pt]
P_{\phi\phi ,22}=
&-\frac{16\mu^6g^2}{(2\pi)^8\epsilon^2}
\int \dif p\,\dif q\,\dif k\,\dif p'\dif q'\dif k'
\delta(p+q+k)\delta(p'\!+q'\!+k')\\
&\times \sin \frac{p\wedge q}{2} 
\sin\frac{p'\!\wedge q'}{2} \,\frac{(p\wedge q)\, (p'\!\wedge q')}{4 (p+q)^2(p'\!+q')^2}\,
K_4(r,s|p,q,k,p'\!,q'\!,k') \nonumber	
\\ 
&-\frac{4g^2}{(2\pi )^6a}
\int \dif p\,\dif q\,\dif p'\dif q'
\delta(p+q+p'\!+q')\nonumber\\
&\times\sin\frac{p\wedge q}{2} \sin\frac{p'\wedge q'}{2}\frac{p\cdot p'}{(p+q)^2}\,
K_3(r,s|p,q,p'\!,q'), 
\nonumber\\[8pt]
P_{\phi\phi,23}=
&\frac{16\mu^6g^2}{(2\pi)^8}
\int \dif p\,\dif q\,\dif k\,\dif p'\dif q'\dif k'
\delta(p+q+k)\delta(p'\!+q'\!+k')\\
&\times\cos\frac{p\wedge q}{2} \sin\frac{p'\!\wedge q'}{2}\,
\frac{(p'\!\wedge q')\,(p+q)^\mu}{2 (p+q)^2(p'\!+q')^2}\,
\frac{\partial}{\partial p^\mu}K_4(r,s|p,q,k,p'\!,q'\!,k') 	
\nonumber\\ 
&-\frac{8\epsilon\mu^2g^2}{(2\pi)^6a}
\int \dif p\,\dif q\,\dif p'\dif q'
\delta(p+q+p'\!+q')\nonumber\\
&\times\cos\frac{p\wedge q}{2} \sin\frac{p'\!\wedge q'}{2}
\frac{\tilde{p}'^\mu}{(p+q)^2}
\,\frac{\partial}{\partial p^\mu}K_3(r,s|p,q,p'\!,q'), 
&\nonumber\\[8pt]
P'_{\phi\phi, 25}=
&\frac{8\mu^6g^2}{(2\pi)^6\epsilon^2} 
\int \dif p\,\dif q\,\dif p'\dif q' 
\delta(p+q+p'\!+q')\\
&\times\sin\frac{p\wedge q}{2} \sin\frac{p'\!\wedge q'}{2}
\, \frac{(p\wedge q)\, (p'\!\wedge q')}{4 (p+q)^2p'^2q'^2}\,
K_3(r,s|p,q,p'\!,q'),\nonumber\\[8pt]
P_{\phi\phi ,33}=
&-\frac{16\epsilon^2\mu^6g^2}{(2\pi)^8} 
\int \dif p\,\dif q\,\dif k\,\dif p'\dif q'\dif k' 
\delta(p+q+k)\delta(p'\!+q'\!+k')\\
&\times\cos\frac{p\wedge q}{2} \cos\frac{p'\!\wedge q'}{2}
\, \frac{(p+q)_\mu {(p'\!+q')}_\nu}{(p+q)^2(p'\!+q')^2} 
\, \frac{\partial^2}{\partial p_\mu \partial p\smash{'}_{\!\!\nu}}K_4(r,s|p,q,k,p'\!,q'\!,k')
\nonumber\\
&-\frac{16\epsilon^2\mu^4g^2}{(2\pi)^6a} 
\int \dif p\,\dif q\,\dif p'\dif q' 
\delta(p+q+p'\!+q')\nonumber\\
&\times\cos\frac{p\wedge q}{2} \cos\frac{p'\!\wedge q'}{2}\frac{1}{(p+q)^2} \,
\frac{\partial^2}{\partial p^\mu \partial p'_\mu}K_3(r,s|p,q,p'\!,q'),\nonumber\\[8pt]
P'_{\phi\phi, 35}=&\frac{8\mu^6g^2}{(2\pi)^6} 
\int \dif p\,\dif q\,\dif p'\dif q' 
\delta(p+q+p'\!+q')\\
&\times\sin\frac{p\wedge q}{2} \cos\frac{p'\!\wedge q'}{2}
\frac{(p\wedge q)\,(p+q)_\mu}{2 p^2q^2(p+q)^2} \,
\frac{\partial}{\partial p'_\mu}K_3(r,s|p,q,p'\!,q')\nonumber\\[8pt]
P'_{\phi\phi,45}=&-\frac{8a\mu^8g^2}{(2\pi)^6\epsilon^2}
\int \dif p\,\dif q\,\dif p'\dif q' 
\delta(p+q+p'\!+q')\\
&\times\sin\frac{p\wedge q}{2} \sin\frac{p'\!\wedge q'}{2} 
\, \frac{(p\wedge q)\,(p'\!\wedge q')}{4 p^2q^2p'^2(p'\!+q')^2}
 \, K_3(r,s|p,q,p'\!,q')
\nonumber \\
&+\frac{2\mu^2g^2}{(2\pi)^4}
\int \dif p\,\dif q\,
\sin^2\frac{p\wedge q}{2} \frac{1}{p^2q^2} \, K_2(r,s|p+q,-p-q)\nonumber\\[8pt]
P'_{\phi\phi,55}=&\frac{8a\mu^8g^2}{(2\pi)^6\epsilon^2}
\int \dif p\,\dif q\,\dif p'\dif q' 
\delta(p+q+p'\!+q')
\sin\frac{p\wedge q}{2} \\
&\times\sin\frac{p'\!\wedge q'}{2}
 \,\frac{(p\wedge q)\,(p'\!\wedge q')}{4p^2q^2p'^2(p'\!+q')^2} \, K_3(r,s|p,q,p'\!,q')\nonumber
\\
&+\frac{4a\mu^8g^2}{(2\pi)^6\epsilon^2}
\int \dif p\,\dif q\,\dif p'\dif q' 
\delta(p+q+p'\!+q')\nonumber \\
&\times\sin\frac{p\wedge q}{2} \sin\frac{p'\!\wedge q'}{2}
\,  \frac{(p\wedge q)\,(p'\!\wedge q')}{4 p^2q^2p'^2q'^2} \, K_3(r,s|p,q,p'\!,q')\nonumber
\\
&+\frac{4\mu^2g^2}{(2\pi)^4}
\int \dif p\,\dif q\,
\sin^2\frac{p\wedge q}{2}\nonumber\\
&\times\left(
-\frac{1}{p^2q^2}
+\frac{p\cdot q}{p^2q^2(p+q)^2} 
-\frac{1}{(p+q)^4}
+\frac{(p \cdot q)^2}{p^2q^2(p+q)^4}
\right) K_2(r,s|p+q,-p-q) 
&\nonumber\\[8pt]
P_{\phi\phi,66}=
&-\frac{4\mu^2 g^2}{(2\pi)^4}
\int \dif p\,\dif q\, 
\sin^2\frac{p\wedge q}{2} \left(\frac{1}{(p+q)^4}-\frac{(p\cdot q)^2}{p^2q^2(p+q)^4}\right)
 K_2(r,s|p+q,-p-q)                 .\label{Pff66}
\end{align}
They correspond to the additional diagrams:
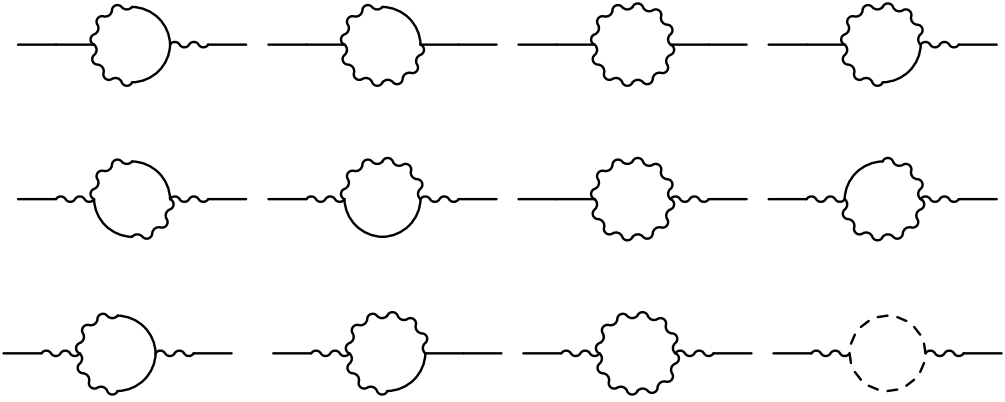
\begin{center}
\begin{fmffile}{propagator2}
 \begin{fmfgraph}(30,20)
 \fmfleft{p,i,m}\fmfright{q,o,n}
 \fmf{phantom,tension=0.5}{p,t}\fmf{phantom,tension=0.5}{t,q}
  \fmf{phantom,tension=0.5}{m,b}\fmf{phantom,tension=0.5}{b,n}
 \fmf{plain,tension=1}{i,v}
 \fmf{plain,tension=1}{v,v1}
 \fmf{wiggly,right=0.5,tension=0.5}{v1,t}
 \fmf{plain,right=0.4,tension=0.5}{t,v2}
 \fmf{plain,right=0.4,tension=0.5}{v2,b}
 \fmf{wiggly,right=0.5,tension=0.5}{b,v1}
 \fmf{wiggly,tension=1}{v2,v3}
 \fmf{plain,tension=1}{v3,o}
 \end{fmfgraph}
 \begin{fmfgraph}(30,20)
 \fmfleft{p,i,m}\fmfright{q,o,n}
 \fmf{phantom,tension=0.5}{p,t}\fmf{phantom,tension=0.5}{t,q}
  \fmf{phantom,tension=0.5}{m,b}\fmf{phantom,tension=0.5}{b,n}
 \fmf{plain,tension=1}{i,v}
 \fmf{plain,tension=1}{v,v1}
 \fmf{wiggly,right=0.5,tension=0.5}{v1,t}
 \fmf{wiggly,right=0.5,tension=0.5}{t,v2}
 \fmf{plain,right=0.4,tension=0.5}{v2,b}
 \fmf{wiggly,right=0.5,tension=0.5}{b,v1}
 \fmf{plain,tension=1}{v2,v3}
 \fmf{plain,tension=1}{v3,o}
 \end{fmfgraph}
\begin{fmfgraph}(30,20)
 \fmfleft{p,i,m}\fmfright{q,o,n}
 \fmf{phantom,tension=0.5}{p,t}\fmf{phantom,tension=0.5}{t,q}
  \fmf{phantom,tension=0.5}{m,b}\fmf{phantom,tension=0.5}{b,n}
 \fmf{plain,tension=1}{i,v}
 \fmf{plain,tension=1}{v,v1}
 \fmf{wiggly,right=0.5,tension=0.5}{v1,t}
 \fmf{wiggly,right=0.5,tension=0.5}{t,v2}
 \fmf{wiggly,right=0.5,tension=0.5}{v2,b}
 \fmf{wiggly,right=0.5,tension=0.5}{b,v1}
 \fmf{plain,tension=1}{v2,v3}
 \fmf{plain,tension=1}{v3,o}
 \end{fmfgraph}
 \begin{fmfgraph}(30,20)
 \fmfleft{p,i,m}\fmfright{q,o,n}
 \fmf{phantom,tension=0.5}{p,t}\fmf{phantom,tension=0.5}{t,q}
  \fmf{phantom,tension=0.5}{m,b}\fmf{phantom,tension=0.5}{b,n}
 \fmf{plain,tension=1}{i,v}
 \fmf{plain,tension=1}{v,v1}
 \fmf{wiggly,right=0.5,tension=0.5}{v1,t}
 \fmf{plain,right=0.4,tension=0.5}{t,v2}
 \fmf{wiggly,right=0.5,tension=0.5}{v2,b}
 \fmf{wiggly,right=0.5,tension=0.5}{b,v1}
 \fmf{wiggly,tension=1}{v2,v3}
 \fmf{plain,tension=1}{v3,o} 
 \end{fmfgraph}\\
\begin{fmfgraph}(30,20)
 \fmfleft{p,i,m}\fmfright{q,o,n}
 \fmf{phantom,tension=0.5}{p,t}\fmf{phantom,tension=0.5}{t,q}
  \fmf{phantom,tension=0.5}{m,b}\fmf{phantom,tension=0.5}{b,n}
 \fmf{plain,tension=1}{i,v}
 \fmf{wiggly,tension=1}{v,v1}
 \fmf{plain,right=0.4,tension=0.5}{v1,t}
 \fmf{wiggly,right=0.5,tension=0.5}{t,v2}
 \fmf{plain,right=0.4,tension=0.5}{v2,b}
 \fmf{wiggly,right=0.5,tension=0.5}{b,v1}
 \fmf{wiggly,tension=1}{v2,v3}
 \fmf{plain,tension=1}{v3,o}
 \end{fmfgraph}
 \begin{fmfgraph}(30,20)
 \fmfleft{p,i,m}\fmfright{q,o,n}
 \fmf{phantom,tension=0.5}{p,t}\fmf{phantom,tension=0.5}{t,q}
  \fmf{phantom,tension=0.5}{m,b}\fmf{phantom,tension=0.5}{b,n}
 \fmf{plain,tension=1}{i,v}
 \fmf{wiggly,tension=1}{v,v1}
 \fmf{plain,right=0.4,tension=0.5}{v1,t}
 \fmf{plain,right=0.4,tension=0.5}{t,v2}
 \fmf{wiggly,right=0.5,tension=0.5}{v2,b}
 \fmf{wiggly,right=0.5,tension=0.5}{b,v1}
 \fmf{wiggly,tension=1}{v2,v3}
 \fmf{plain,tension=1}{v3,o}
 \end{fmfgraph}
 \begin{fmfgraph}(30,20)
 \fmfleft{p,i,m}\fmfright{q,o,n}
 \fmf{phantom,tension=0.5}{p,t}\fmf{phantom,tension=0.5}{t,q}
  \fmf{phantom,tension=0.5}{m,b}\fmf{phantom,tension=0.5}{b,n}
 \fmf{plain,tension=1}{i,v}
 \fmf{plain,tension=1}{v,v1}
 \fmf{wiggly,right=0.5,tension=0.5}{v1,t}
 \fmf{wiggly,right=0.5,tension=0.5}{t,v2}
 \fmf{wiggly,right=0.5,tension=0.5}{v2,b}
 \fmf{wiggly,right=0.5,tension=0.5}{b,v1}
 \fmf{wiggly,tension=1}{v2,v3}
 \fmf{plain,tension=1}{v3,o}
 \end{fmfgraph}
 \begin{fmfgraph}(30,20)
 \fmfleft{p,i,m}\fmfright{q,o,n}
 \fmf{phantom,tension=0.5}{p,t}\fmf{phantom,tension=0.5}{t,q}
  \fmf{phantom,tension=0.5}{m,b}\fmf{phantom,tension=0.5}{b,n}
 \fmf{plain,tension=1}{i,v}
 \fmf{wiggly,tension=1}{v,v1}
 \fmf{wiggly,right=0.5,tension=0.5}{v1,t}
 \fmf{wiggly,right=0.5,tension=0.5}{t,v2}
 \fmf{wiggly,right=0.5,tension=0.5}{v2,b}
 \fmf{plain,right=0.4,tension=0.5}{b,v1}
 \fmf{wiggly,tension=1}{v2,v3}
 \fmf{plain,tension=1}{v3,o}
 \end{fmfgraph}\\
 \begin{fmfgraph}(30,20)
 \,\,\,\,\fmfleft{p,i,m}\fmfright{q,o,n}
 \fmf{phantom,tension=0.5}{p,t}\fmf{phantom,tension=0.5}{t,q}
  \fmf{phantom,tension=0.5}{m,b}\fmf{phantom,tension=0.5}{b,n}
 \fmf{plain,tension=1}{i,v}
 \fmf{wiggly,tension=1}{v,v1}
 \fmf{wiggly,right=0.5,tension=0.5}{v1,t}
 \fmf{plain,right=0.4,tension=0.5}{t,v2}
 \fmf{plain,right=0.4,tension=0.5}{v2,b}
 \fmf{wiggly,right=0.5,tension=0.5}{b,v1}
 \fmf{wiggly,tension=1}{v2,v3}
 \fmf{plain,tension=1}{v3,o}
 \end{fmfgraph}
\begin{fmfgraph}(30,20)
 \fmfleft{p,i,m}\fmfright{q,o,n}
 \fmf{phantom,tension=0.5}{p,t}\fmf{phantom,tension=0.5}{t,q}
  \fmf{phantom,tension=0.5}{m,b}\fmf{phantom,tension=0.5}{b,n}
 \fmf{plain,tension=1}{i,v}
 \fmf{wiggly,tension=1}{v,v1}
 \fmf{wiggly,right=0.5,tension=0.5}{v1,t}
 \fmf{plain,right=0.4,tension=0.5}{t,v2}
 \fmf{wiggly,right=0.5,tension=0.5}{v2,b}
 \fmf{wiggly,right=0.5,tension=0.5}{b,v1}
 \fmf{plain,tension=1}{v2,v3}
 \fmf{plain,tension=1}{v3,o}
 \end{fmfgraph}
 \begin{fmfgraph}(30,20)
 \fmfleft{p,i,m}\fmfright{q,o,n}
 \fmf{phantom,tension=0.5}{p,t}\fmf{phantom,tension=0.5}{t,q}
  \fmf{phantom,tension=0.5}{m,b}\fmf{phantom,tension=0.5}{b,n}
 \fmf{plain,tension=1}{i,v}
 \fmf{wiggly,tension=1}{v,v1}
 \fmf{wiggly,right=0.5,tension=0.5}{v1,t}
 \fmf{wiggly,right=0.5,tension=0.5}{t,v2}
 \fmf{wiggly,right=0.5,tension=0.5}{v2,b}
 \fmf{wiggly,right=0.5,tension=0.5}{b,v1}
 \fmf{wiggly,tension=1}{v2,v3}
 \fmf{plain,tension=1}{v3,o}
 \end{fmfgraph}
\begin{fmfgraph}(30,20)
 \fmfleft{p,i,m}\fmfright{q,o,n}
 \fmf{phantom,tension=0.5}{p,t}\fmf{phantom,tension=0.5}{t,q}
  \fmf{phantom,tension=0.5}{m,b}\fmf{phantom,tension=0.5}{b,n}
 \fmf{plain,tension=1}{i,v}
 \fmf{wiggly,tension=1}{v,v1}
 \fmf{dashes,right=0.4,tension=0.5}{v1,t}
 \fmf{dashes,right=0.4,tension=0.5}{t,v2}
 \fmf{dashes,right=0.4,tension=0.5}{v2,b}
 \fmf{dashes,right=0.4,tension=0.5}{b,v1}
 \fmf{wiggly,tension=1}{v2,v3}
 \fmf{plain,tension=1}{v3,o}
 \end{fmfgraph}
\end{fmffile}
\end{center}

The dashed line is the ghost propagator 
$\bcontraction[0.5ex]{}{\bar{c}}{}{c}\bar{c}c$. The one-loop
propagator corrections  $P_{\phi\phi,7}$,
 $P_{\phi\phi ,8}$, $P_{\phi\phi, 9}$ and $P_{\phi\phi ,10}$ which were calculated in 
 \cite{Buric:2012bb} are
\begin{align}
P_{\phi\phi,7}=
&\frac{2\mu^2g^2}{(2\pi)^6}
\int \dif p\,\dif q\,\dif p'\dif q' 
\delta(p+q+p'\!+q')
\sin\frac{p\wedge q}{2} \sin\frac{p'\!\wedge q'}{2} \frac{p \cdot p'}{p^2p'^2}
\, K_3(r,s|p,q,p'\!,q')
\nonumber \\ 
&+\frac{4g^2}{(2\pi)^4a}
\int \dif p\,\dif q\,
\sin^2\frac{p\wedge q}{2}
\frac{1}{p^2} \, K_2(r,s|q,-q)
\nonumber\\[8pt]
P_{\phi\phi,8}=
&\frac{2\epsilon^2\mu^2g^2}{(2\pi)^6}
\int \dif p\,\dif q\,\dif p'\dif q' 
\delta(p+q+p'\!+q')
\cos\frac{p\wedge q}{2} \cos\frac{p'\!\wedge q'}{2} \frac{p \cdot p'}{p^2p'^2} 
\, K_3(r,s|p,q,p'\!,q')
\nonumber \\
&+\frac{4\epsilon^2g^2}{(2\pi)^4a}
\int \dif p\,\dif q\,
\cos^2\frac{p\wedge q}{2}
\frac{1}{p^2} \, K_2(r,s|q,-q)\\[8pt]
P_{\phi\phi,9}=
&-\frac{4\mu^4g^2}{(2\pi)^6}
\int \dif p\,\dif q\,\dif p'\dif q' 
\delta(p+q+p'\!+q')\\
&\times\sin\frac{p\wedge q}{2} \cos\frac{p'\!\wedge q'}{2} 
\, \frac{p\wedge q}{2 p^2q^2} \, K_3(r,s|p,q,p'\!,q')
\nonumber\\[8pt]
P_{\phi\phi,10}=
&-\frac{2a\mu^8g^2}{(2\pi)^6\epsilon^2}
\int \dif p\,\dif q\,\dif p'\dif q' 
\delta(p+q+p'\!+q')\\
&\times\sin\frac{p\wedge q}{2} \sin\frac{p'\!\wedge q'}{2} \,
\frac{(p\wedge q)\,(p'\!\wedge q')}{4 p^2q^2p'^2q'^2} \, K_3(r,s|p,q,p'\!,q')
\nonumber \\
&+\frac{2\mu^2g^2}{(2\pi)^4}
\int \dif p\,\dif q\,
\sin^2\frac{p\wedge q}{2}
\left(
\frac{1}{2p^2q^2}-\frac{p \cdot q}{p^2q^2(p+q)^2} 
\right)
\, K_2(r,s|p+q,-p-q)\nonumber .
\end{align}
The ghost contributions are zero.

\vskip1cm
\noindent
\begin{large}
{\bf Appendix 2}
\end{large}
\vskip0.5cm

Some of the Gaussian integrals in two dimensions:
\begin{eqnarray*}
&& \int dp \, e^{-ap^2+b\cdot p}=\frac{\pi}{a}\,e^{b^2/4a}
\\[8pt] &&
\int dp\,p_\alpha p_\beta \, e^{-ap^2+b\cdot p}
=\frac{\pi}{2a^2}\left(\delta_{\alpha\beta}
+\frac{b_\alpha b_\beta}{2a}\right)\,e^{b^2/4a}
\\[8pt] &&
\int dp\,(p^2)^2 \, e^{-ap^2+b\cdot p}=\frac{\pi}{a^3}
\left(2+\frac{b^2}{a}+\frac{(b^2)^2}{16a^2}\right)\,e^{b^2/4a}
\\[8pt] &&
\int dp \, \frac{p_\alpha p_\beta}{p^2} \, e^{-ap^2+b\cdot p}=
\frac{2\pi}{b^2}\left(\frac{b^2\delta_{\alpha\beta}-2b_\alpha b_\beta}{b^2}
+\frac{b_\alpha b_\beta}{2a}\right)\,e^{b^2/4a}
\end{eqnarray*}

Switching from $u$-integration to $u^2$-integration:
\begin{eqnarray*}
 &&\int du \, f(u^2) \, \frac{u_\alpha u_\beta}{u^2} 
 =\frac{\pi}{2}\delta_{\alpha\beta} \int d(u^2) \, f(u^2)
\\[8pt] &&
\int du \, f(u^2) \, \frac{u_\alpha u_\beta u_\gamma u_\delta}{(u^2)^2} 
 = \frac{\pi}{8}(\delta_{\alpha\beta}\delta_{\gamma\delta}
 +\delta_{\alpha\gamma}\delta_{\beta\delta}+\delta_{\alpha\delta}
 \delta_{\beta\gamma}) \int d(u^2) \, f(u^2)
\end{eqnarray*}

Some formulas used to evaluate the amputated propagators:
\begin{align}
&\frac{1}{(2\pi)^4}\int du\, K_2(p,q,k,u)K^{-1}(-u,r)=
\nonumber\\
&\qquad =\delta(p+r)K(q,k)+\delta(q+r)K(p,q)
+\delta(k+r)K(p,q)
\nonumber\\[8pt]
&\frac{1}{(2\pi)^8}\int du\,dv\, K^{-1}(r,-u)K_3(u,v|p,q,p'
\!,q')K^{-1}(-u,r)=\nonumber \\
&\qquad = \delta(p+r)\delta(q+s)K(p'\!,q')
+\delta(p+r)\delta(p'\!+s)K(q,q')+\delta(p+r)\delta(q'\!+s)
K(q,p') \nonumber \\[6pt]
&\qquad +\delta(q+r)\delta(p+s)K(p'\!,q')+\delta(q+r)\delta(q'\!+s)K(p,p')
+\delta(q+r)\delta(p'\!+s)K(p,q) \nonumber \\[6pt]
&\qquad + (r\leftrightarrow s )\nonumber.
\end{align}

\vskip1cm
\noindent
\begin{large}
{\bf Appendix 3}
\end{large}
\vskip0.5cm

Divergent contributions to the $\phi\phi$-part 
of the effective action after expansion around $u=0$ are
\begin{align}
\Gamma^{(2)}_{\phi\phi}&=-\frac{3\mu^2 ag^2}{\epsilon^2}
\int du\,dv\,\frac{\phi(-v)\phi(v)}{(v^2)^2}
\int\limits^{\infty}_{1}\frac{d\xi}{\xi}\frac{\xi-1}{\xi+1} \, e^{-\xi\frac{u^2}{2\mu^2}}
\int\limits^{\infty}_{0}d\eta\, 
e^{-\frac{\xi\epsilon^2 }{1+\eta\xi}\frac{v^2}{2\mu^2}}
\label{gama2}
\\
&-a g^2
\int du\,dv\,\frac{\phi(-v)\phi(v)}{v^2}
\int\limits^{\infty}_{1}d\xi\,\frac{\xi-1}{\xi+1} \, e^{-\xi\frac{u^2}{2\mu^2}}
\int\limits^{\infty}_{0}\frac{d\eta}{1+\eta\xi}\, 
e^{-\frac{\xi\epsilon^2}{1+\eta\xi}\frac{v^2}{2\mu^2}}
\nonumber
\\
&+\frac{a g^2}{2}
\int du\,dv\,\frac{\phi(-v)\phi(v)}{v^2}
\int\limits^{\infty}_{1}d\xi\,\frac{\xi-1}{\xi+1} \, e^{-\xi\frac{u^2}{2\mu^2}}
\int\limits^{\infty}_{0}\frac{d\eta}{(1+2\eta\xi)^2-\epsilon^2\xi^2}
\nonumber
\\
&\hspace{160pt}\times\Bigg(2-\frac{1+\epsilon^2\xi^2}{1+\eta\xi}-\frac{2\epsilon^2\xi^3}{(1+2\eta\xi)^2-\epsilon^2\xi^2}\frac{\epsilon^2\xi+\eta}{1+\eta\xi}\Bigg)
\nonumber
\end{align}

\begin{align}
\Gamma^{(3)}_{\phi\phi}&=
\frac{\mu^2 g^2}{\epsilon^2}
\int du\,dv\, \frac{\phi(-v)\phi(v)}{(v^2)^2}
\int\limits^{\infty}_{1}\frac{d\xi}{\xi}\frac{\xi-1}{\xi+1} \, e^{-\xi\frac{u^2}{2\mu^2}}
\int\limits^{\infty}_{0}d\eta\, 
e^{-\frac{\xi\epsilon^2 }{1+\eta\xi}\frac{v^2}{2\mu^2}}
\label{gama3}
\\
&+\mu^2 g^2
\int du\,dv\, \frac{\phi(-v)\phi(v)}{(v^2)^2}
\int\limits^{\infty}_{1}d\xi\,\xi\,\frac{\xi-1}{\xi+1} \, e^{-\xi\frac{u^2}{2\mu^2}}
\int\limits^{\infty}_{0}d\eta\,\frac{(1+2\eta\xi)^2+\epsilon^2\xi^2}
{((1+2\eta\xi)^2-\epsilon^2\xi^2)^2} \nonumber
\\
&-2\epsilon^2\mu^2 g^2
\int \frac{du}{u^2}\,dv\, \frac{\phi(-v)\phi(v)}{v^2}
\int\limits^{\infty}_{1}d\xi\,\xi^3\,\frac{\xi-1}{\xi+1} \, e^{-\xi\frac{u^2}{2\mu^2}}
\int\limits^{\infty}_{0}\frac{d\eta}{((1+2\eta\xi)^2-\epsilon^2\xi^2)^2}
\nonumber
\\
&-g^2
\int du\,dv\, \frac{\phi(-v)\phi(v)}{v^2}
\int\limits^{\infty}_{1}d\xi\,\frac{\xi-1}{\xi+1} \, e^{-\xi\frac{u^2}{2\mu^2}}
\int\limits^{\infty}_{0}d\eta\,\frac{(1+2\eta\xi)^2}{(1+2\eta\xi)^2-\epsilon^2\xi^2}
\nonumber
\\
&+\frac{g^2}{2}
\int du\,dv\, \frac{\phi(-v)\phi(v)}{v^2}
\int\limits^{\infty}_{1}d\xi\,\frac{\xi-1}{\xi+1} \, e^{-\xi\frac{u^2}{2\mu^2}}
\int\limits^{\infty}_{0}\frac{d\eta}{1+\eta\xi}
\frac{(1+2\eta\xi)^2+\epsilon^2\xi^2}{(1+2\eta\xi)^2-\epsilon^2\xi^2}
\nonumber
\\
&+\epsilon^2 g^2
\int du\,dv\, \frac{\phi(-v)\phi(v)}{v^2}
\int\limits^{\infty}_{1}d\xi\,\xi^3\,\frac{\xi-1}{\xi+1} \, e^{-\xi\frac{u^2}{2\mu^2}}
\int\limits^{\infty}_{0}\frac{d\eta}{1+\eta\xi}
\frac{\epsilon^2\xi+\eta}{((1+2\eta\xi)^2-\epsilon^2\xi^2)^2}
\nonumber
\end{align}

\begin{align}
\Gamma^{(4)}_{\phi\phi}&=a\mu^2 g^2
\int du\,dv\, \frac{\phi(-v)\phi(v)}{(v^2)^2}
\int\limits^{\infty}_{1}\frac{d\xi}{\xi}\frac{\xi-1}{\xi+1} \, e^{-\xi\frac{u^2}{2\mu^2}}
\int\limits^{\infty}_{0}\frac{d\eta}{\eta^2}\, 
e^{-\frac{\eta}{1+\eta\xi}\frac{v^2}{2\mu^2}}
\label{gama4}
\\
&-a\mu^2 g^2
\int du\,dv\, \frac{\phi(-v)\phi(v)}{(v^2)^2}
\int\limits^{\infty}_{1}\frac{d\xi}{\xi}\frac{\xi-1}{\xi+1} \, e^{-\xi\frac{u^2}{2\mu^2}}
\int\limits^{\infty}_{0}d\eta\,\frac{\eta^2+\epsilon^2}{(\eta^2-\epsilon^2)^2}\, 
e^{-\frac{\eta+\xi\epsilon^2}{1+\eta\xi}\frac{v^2}{2\mu^2}}
\nonumber
\\
&+a\epsilon^2 g^2
\int du\,dv\, \frac{\phi(-v)\phi(v)}{v^2}
\int\limits^{\infty}_{1}d\xi\,\frac{\xi-1}{\xi+1} \, e^{-\xi\frac{u^2}{2\mu^2}}
\int\limits^{\infty}_{0}\frac{d\eta}{(\eta^2-\epsilon^2)(1+\eta\xi)}\, 
e^{-\frac{\eta+\xi\epsilon^2}{1+\eta\xi}\frac{v^2}{2\mu^2}}
\nonumber
\end{align}


\newpage

\noindent
\begin{large}
{\bf Appendix 4}
\end{large}
\vskip0.5cm

\noindent
{\bf 4.1 Relation between the $\xi$-integral and the $u^2$-divergence}
\vskip0.5cm

As  seen in the  formulas throughout the paper, the integrals 
over $\xi$ contain an exponent  $\lambda=u^2/2\mu^2$
 which regularizes them at the upper bound
\begin{equation}
\int\limits_1^\infty d\xi\,f(\xi)\,e^{-\lambda\xi} .
\end{equation}
This exponent is lost if we expand in small $u^2$ by putting
$u^2=0$: in that case the IR divergence in $u$
 is transferred to  divergence in $\xi$.
For better control, we will keep the exponent.

The encountered integrals can be written as 
\begin{equation}
\int\limits_1^\infty d\xi\,f(\xi)\, e^{-\lambda\xi}
=\int\limits_1^\infty d\xi\,g(\xi)\, h(1/\xi)\, e^{-\lambda\xi},
\end{equation}
where $g(\xi)$ is a simpler function of $\xi$ and $h(\xi)$ 
has the form
\begin{equation}
h(\xi)=H+O(1/\xi),                          \label{H}
\qquad
H={\rm const.}
\end{equation}
Let us analyze two integrals
\begin{equation}
I_1(\lambda)=\int\limits_1^\infty d\xi\,g(\xi)\, e^{-\lambda\xi},
\qquad
I_2(\lambda)=\int\limits_1^\infty d\xi\,\frac{g(\xi)}{\xi}\, e^{-\lambda\xi},
\qquad
\frac{dI_2}{d\lambda}=-I_1.
\end{equation}
The last equation implies that if $\, I_2=O(\lambda^n)$ when 
$\lambda\rightarrow0$, then $\, I_1=O(\lambda^{n-1})$ and
\begin{equation}
\frac{I_2}{I_1}\rightarrow0,\qquad \lambda\rightarrow0.
\end{equation}
In other words if both $I_1$ and $I_2$ are divergent, 
then the $I_1$-divergence is of the higher order.
Since we are interested in the leading divergence in $\lambda$, we 
can discard the $O(1/\xi)$ contribution in (\ref{H})
and simplify the initial integral:
\begin{equation}
\int\limits_1^\infty d\xi\,f(\xi)\,e^{-\lambda\xi} \longrightarrow
H\int\limits_1^\infty d\xi\,g(\xi)\,e^{-\lambda\xi}.
\label{simplification}
\end{equation}

The actual simplified integrals appearing in  the effective 
action are the following  ($n>0$, $\gamma$ is the Euler-Mascheroni constant): 
\begin{eqnarray}
&& \int\limits_1^\infty d\xi\,\xi^n e^{-\lambda\xi}=
\dfrac{n!}{\lambda^{n+1}}-\dfrac{1}{n+1}+O(\lambda),
\nonumber\\[4pt]
&&\int\limits_1^\infty \dfrac{d\xi}{\xi} e^{-\lambda\xi}=
-\log\lambda-\gamma+O(\lambda),
\nonumber\\[4pt] 
&& \int\limits_1^\infty \dfrac{d\xi}{\xi^{n+1}} e^{-\lambda\xi}
= \dfrac{1}{n}+O(\lambda),
\nonumber\\[4pt]
&& \int\limits_1^\infty d\xi\,\dfrac{\log\xi}{\xi} e^{-\lambda\xi}=
\dfrac{1}{2}\log^2\lambda+\gamma\log\lambda+\frac{\gamma^2}{2}+\dfrac{\pi^2}{12}+
O(\lambda).\nonumber
\end{eqnarray}

\vskip0.5cm
\noindent
{\bf 4.2 Detailed analysis}
\vskip0.5cm

Expansion around $u=0$ gives, besides (\ref{454}), further 
 terms listed in Appendix 3 which are potentially divergent at the
 lower boundary of integration in $u$. The selected terms contain 
 \begin{equation}
 \int dv\, \frac{\phi(-v)\phi(v)}{(v^2)^2},\qquad
 \int dv\, \frac{\phi(-v)\phi(v)}{v^2}
 \end{equation}
 which give nonlocal contributions to the one-loop effective
 action, 
 \begin{equation}
 \int\phi\,\Box^{-2}\phi, \qquad \int\phi\,\Box^{-1}\phi.
 \end{equation}
 All expressions contain integrations over two parameters:
 we wish to sum divergent contributions and see weather the result is 
 zero, finite or divergent. We first observe the exponentials in $v^2$
 in $\, \Gamma^{(2)}_{\phi\phi}$ and $\, \Gamma^{(4)}_{\phi\phi}$. In order 
 to extract the $\Box^{-1}$- and $\Box^{-2}$-parts
 of the one-loop effective action, we expand this
exponential in power series and consider only the first two terms: 
the remaining ones give local contributions. 

To explain the regularization procedure, we start with the integral
\begin{equation}
I=\int du\,dv\, \frac{\phi(-v)\phi(v)}{(v^2)^2}
\int\limits^{\infty}_{1}\frac{d\xi}{\xi}\frac{\xi-1}{\xi+1}\,
e^{-\frac{\xi u^2}{2\mu^2}}
\int\limits^{\infty}_{0}d\eta\,
e^{-\frac{\epsilon^2\xi}{1+\eta\xi}\frac{v^2}{2\mu^2}} . \label{I1}
\end{equation}
We introduce regularizations in the $u$- 
and in the $\eta$-integrals. We choose the regulators 
to be defined by the same large parameter $\Lambda$:
$$
\int du\, f(u^2) \longrightarrow \pi\int\limits_{\mu^2/\Lambda} d(u^2)\, f(u^2),
\qquad
\int d\eta \longrightarrow \int\limits_{\beta_2/\Lambda}^{\beta\Lambda} d\eta.
$$
We find
\begin{equation}
I^{(div)}=\pi\int dv \,\frac{\phi(-v)\phi(v)}{(v^2)^2}
\int\limits_{\mu^2/\Lambda} d(u^2) 
\int\limits^{\infty}_{1}\frac{d\xi}{\xi}\frac{\xi-1}{\xi+1}\, 
e^{-\frac{\xi u^2}{2\mu^2}}
\int\limits_{\beta_2/\Lambda}^{\beta\Lambda}d\eta\,
\Bigg(1-\frac{\epsilon^2\xi}{1+\eta\xi}\frac{v^2}{2\mu^2}\Bigg). \label{I1a}
\end{equation}
We can set the lower boundary of the integral over $\eta$ to zero
since it contains no divergence. Integration over $\eta$ and Fourier 
transformation give
\begin{align}
I^{(div)}=4\pi^3
\Bigg(
\beta\Lambda
&\int \phi\,\Box^{-2}\phi
\int\limits_{\mu^2/\Lambda} d(u^2) 
\int\limits^\infty_{1}\frac{d\xi}{\xi}\frac{\xi-1}{\xi+1}\,
e^{-\frac{\xi u^2}{2\mu^2}}
\nonumber
\\ \qquad
-\frac{\epsilon^2}{2\mu^2}
&\int \phi\,\Box^{-1}\phi
\int\limits_{\mu^2/\Lambda} d(u^2)
\int\limits^\infty_{1}\frac{d\xi}{\xi}
\frac{\xi-1}{\xi+1}\log(\beta\Lambda\xi+1)\,e^{-\frac{\xi u^2}{2\mu^2}}
\Bigg).
\nonumber
\end{align}
In accordance with the previous discussion leading to (\ref{simplification}), we keep
only the leading contribution at   $\xi\rightarrow\infty$,
and obtain
\begin{align}
I^{(div)}=4\pi^3
\Bigg(
\beta\Lambda
&\int \phi\,\Box^{-2}\phi
\int\limits_{\mu^2/\Lambda} d(u^2) 
\int\limits^\infty_{1}\frac{d\xi}{\xi}\, e^{-\frac{\xi u^2}{2\mu^2}}
\nonumber
\\ \qquad
-\frac{\epsilon^2}{2\mu^2}
&\int \phi\,\Box^{-1}\phi
\int\limits_{\mu^2/\Lambda} d(u^2)
\int\limits^\infty_{1}\frac{d\xi}{\xi}(\log \Lambda + \log \xi)\, 
e^{-\frac{\xi u^2}{2\mu^2}} \Bigg) .
\nonumber
\end{align}
The remaining integration gives the following leading contributions in $\Lambda$
\begin{align}
 I^{(div)}
&=4\pi^3\mu^2 \Bigg( -\beta\log \Lambda\int \phi\,\Box^{-2}\phi
+\frac{3\epsilon^2}{4\mu^2}\,\frac{\log^2\Lambda}{\Lambda}
 \int \phi\,\Box^{-1}\phi \Bigg).
\label{I1d}
\end{align}
The second term vanishes for $\Lambda\to\infty$.

Inspecting the other terms given in Appendix 3 we see that
some of the integrals have singular points which are
 inside the integration domain. In such cases 
 the regulators are introduced in the following manner. 
 Let $\, \zeta=\xi,\eta\, $ denote the integration parameter and 
$\zeta = \zeta_0$ the pole of the integration function. We regularize 
as before using the same regulator $\Lambda$, replacing
$$ 
\int d\zeta \to \int\limits^{\zeta_0-\frac{\alpha}{\Lambda}} 
d\zeta + \int\limits_{\zeta_0+\frac{\alpha}{\Lambda}} d\zeta .
$$
where $\alpha$ is a positive constant.
Concretely, we make the following substitutions

-- for $\zeta=\eta$ and the pole arrising from $\eta^2-\epsilon^2=0$, 
we denote $\alpha =\gamma$,

-- for $\zeta=\eta$ and the pole arrising from $(1+2\eta\xi)^2
-\epsilon^2\xi^2=0$,\, $\alpha =\gamma_2$,

-- for $\zeta =\xi$ and the pole arrising from 
$\, \log|(1-\epsilon\xi)/(1+\epsilon\xi)|=\infty$, \, $\alpha =\delta$.

\vskip0.5cm
We regularize all integrals given in Appendix 3.
Calculating them  and adding different contributions, 
we obtain that the leading order propagator 
divergences of the one-loop effective action are
\begin{equation}
-4\pi^3\mu^4\left(\frac{1-3a}{\epsilon^2}\beta+\frac{a}{\beta_2}
-\frac{a}{\gamma}+\frac{1+a}{4\gamma_2}\right)\log\Lambda\int 
\phi\,\Box^{-2}\phi
\label{888}
\end{equation}
and
\begin{equation}
-\frac{\epsilon^2\pi^3\mu^2}{2\gamma_2}\Lambda\log^2\Lambda\int \ 
\phi\,\Box^{-1}\phi.
\label{777}
\end{equation}
An additional divergent $\Box^{-1}$-term
comes from the gauge vertices with no Mehler kernel.
It is quadratic in $\Lambda$ and equal to
\begin{equation}
\frac{\epsilon^2\pi^3\mu^2}{\beta_2^2}\Lambda^2\int \phi\,\Box^{-1}\phi .
\label{666}
\end{equation}

Our results contain yet undefined parameters $\beta$, $\beta_2$, 
$\gamma$, $\gamma_2$.They were introduced  to 
examine the possibility to cancel divergences by an appropriate choice of the
regulators. This method  of course is a kind of fine tuning, since one really
does not wish to introduce a large number of different regulators. We however 
find that in any case it is impossible to remove the $\Box^{-1}$ divergence: 
since $\beta_2\neq\infty$,  divergence in (\ref{666}) always remains.
On the other hand, divergent $\Box^{-2}$ term (\ref{888})
can be removed for some values of $a$ by an appropriate choice of $\beta$, $\beta_2$,
 $\gamma$, $\gamma_2$; however, in the  nonpropagating case $a=0$, the term remains.
In the light of this we shall set the parameters to the simplest  value 
$$
\beta=\beta_2=\gamma=\gamma_2=1,
$$
with which the leading one-loop $\, \phi\phi$-propagator divergences become
$$
\left(\frac{8}{\epsilon^2}-14+\epsilon^2\right)\pi^3\mu^4 g^2\log\Lambda\int  \phi\,\Box^{-2}\phi,
$$
and
$$
{\epsilon^2\pi^3\mu^2 g^2}\Lambda^2\int  \phi\,\Box^{-1}\phi .
$$


\end{document}